\documentclass[a4paper, amsfonts, amssymb, amsmath, reprint, showkeys, nofootinbib, twoside, superscriptaddress]{revtex4-2}

\usepackage[english]{babel}
\usepackage[utf8]{inputenc}
\usepackage[colorinlistoftodos, color=green!40, prependcaption]{todonotes}
\usepackage[pdftex, pdftitle={Article}, pdfauthor={Author}]{hyperref} 
\usepackage{color} 
\usepackage{amsmath,amsthm,amssymb}
\usepackage{mathtools}
\usepackage{physics}
\usepackage{xcolor}
\usepackage{graphicx}
\usepackage[left=23mm,right=13mm,top=35mm,columnsep=15pt]{geometry} 
\usepackage{adjustbox}
\usepackage{placeins}
\usepackage[T1]{fontenc}
\usepackage{lipsum}
\usepackage{csquotes}
\usepackage{siunitx}
\usepackage{gensymb}
\usepackage{comment}
\usepackage{cleveref}
\usepackage{longtable}
\usepackage{multirow}
\usepackage{dcolumn}

\DeclareSIUnit\gauss{G}

\begin{document}
\title{Glassy dynamics in a disordered Heisenberg quantum spin system}

\author{A. Signoles*}
\affiliation{Physikalisches Institut, Universit\"at Heidelberg, Im Neuenheimer Feld 226, 69120 Heidelberg, Germany.}
\affiliation{Laboratoire Charles Fabry, Institut d'Optique Graduate School, CNRS, Universit{\'e} Paris-Saclay, 91127 Palaiseau cedex, France}

\author{T. Franz*}
\affiliation{Physikalisches Institut, Universit\"at Heidelberg, Im Neuenheimer Feld 226, 69120 Heidelberg, Germany.}

\author{R. Ferracini Alves}
\affiliation{Physikalisches Institut, Universit\"at Heidelberg, Im Neuenheimer Feld 226, 69120 Heidelberg, Germany.}

\author{M. G\"arttner}
\affiliation{Kirchhoff-Institut f\"ur Physik, Universit\"at Heidelberg, Im Neuenheimer Feld 227, 69120 Heidelberg, Germany}

\author{S. Whitlock}
\affiliation{Physikalisches Institut, Universit\"at Heidelberg, Im Neuenheimer Feld 226, 69120 Heidelberg, Germany.}
\affiliation{IPCMS (UMR  7504) and ISIS (UMR  7006), University of Strasbourg and CNRS, 67000 Strasbourg, France}

\author{G. Z\"urn}
\affiliation{Physikalisches Institut, Universit\"at Heidelberg, Im Neuenheimer Feld 226, 69120 Heidelberg, Germany.}

\author{M. Weidem\"uller}
\affiliation{Physikalisches Institut, Universit\"at Heidelberg, Im Neuenheimer Feld 226, 69120 Heidelberg, Germany.}
\affiliation{Hefei National Laboratory for Physical Sciences at the Microscale and Department of Modern Physics, and CAS Center for Excellence and Synergetic Innovation Center in Quantum Information and Quantum Physics, University of Science and Technology of China, Hefei, Anhui 230026, China.}

\date{\today}

\begin{abstract}
	Understanding the dynamics of strongly interacting disordered quantum systems is one of the most challenging problems in modern science, due to features such as the breakdown of thermalization and the emergence of glassy phases of matter. We report on the observation of anomalous relaxation dynamics in an isolated XXZ quantum spin system realized by an ultracold gas of atoms initially prepared in a superposition of two-different Rydberg states. The total magnetization is found to exhibit sub-exponential relaxation analogous to classical glassy dynamics, but in the quantum case this relaxation originates from the build-up of non-classical correlations. In both experiment and semi-classical simulations, we find the evolution towards a randomized state is independent of the strength of disorder up to a critical value. This hints towards a unifying description of relaxation dynamics in disordered isolated quantum systems, analogous to the generalization of statistical mechanics to out-of-equilibrium scenarios in classical spin glasses.
\end{abstract}

\maketitle

\section{Introduction}

The far-from-equilibrium behavior of isolated quantum systems and in particular their relaxation towards equilibrium still evades a unifying description. It has been conjectured that these systems generically relax to a state of local thermal equilibrium according to the eigenstate thermalization hypothesis (ETH)~\cite{Rigol2008}. However, the ETH does not explain how the equilibrium state will be reached, or even if it will be reached in experimentally accessible timescales. Particularly rich relaxation dynamics are found in disordered quantum systems, where the interplay between interactions and randomness can give rise to new and intrinsically non-equilibrium effects such as pre-thermalization~\cite{Prufer2018,Eigen2018,Erne2018}, many-body localization~\cite{Yao2014,Smith2016,Choi2016}, Floquet time crystals~\cite{Choi2017_tc, Zhang2017}, and quantum scars~\cite{Bernien2017, Turner2018}.

In contrast, most natural systems (e.g., in condensed matter) are not fully isolated from their environment and hence always relax to thermal equilibrium imposed by the external bath~\cite{Parameswaran2017}. But it is known that disorder and frustration effects can lead to a dramatic slowdown of thermalization, associated with the onset of glassy behavior~\cite{Binder1986}. A key signature of this behavior is that macroscopic observables relax in a characteristically non-exponential way, as encountered for example, in doped semiconductors~\cite{Harris2018} and organic superconductors~\cite{Gezo2013}, quasi crystals~\cite{Dzugutov1995}, atoms in optical lattices~\cite{Luschen2017} or diamond color centers~\cite{Choi2017,Kucsko2018}. This raises the question whether slow relaxation, which appears to be ubiquitous in open disordered systems, also emerges in isolated quantum systems. 

A prototypical model for studying far-from-equilibrium quantum dynamics is the Heisenberg XXZ Hamiltonian for spin-1/2 particles. Compared to the Ising Hamiltonian, this class of spin systems has fewer conserved quantities and shows complex, chaotic far-from-equilibrium dynamics which are difficult to describe theoretically~\cite{Hazzard2013}. Here, we experimentally realize a disordered quantum Heisenberg-XXZ spin-1/2 model in an ultracold atomic gas by encoding the spin degree of freedom in two electronically excited (Rydberg) states of each atom. Spin-spin interactions arise naturally through the state-dependent dipolar interactions between Rydberg states, while disorder originates from the random positions of each atom in the gas which gives rise to distance dependent couplings~\cite{Orioli2018}. Using a strong microwave field pulse that couples the two Rydberg states, we initialize the spins in a far-from-equilibrium state and probe their time evolution, thus employing our system as a quantum simulator for unitary spin dynamics in a disordered system out of equilibrium~\cite{Simon2011,Yan2013, Labuhn2016, Bernien2017, Lepoutre2019, Patscheider2019}.

In a large ensemble of Rydberg spins, we observe that the magnetisation follows a sub-exponential dependence characterised by a stretching exponent that is independent of the strength of disorder up to a critical value. Our experiments and supporting numerical simulations suggest that such glassy dynamics, commonly known in disordered open systems, might also be a generic feature of isolated quantum spin systems, hinting towards a unifying effective theory description.

In Sec.~\ref{sec:qualitative_picture}, we give a qualitative physical picture by solving the time-dependent Schr\"odinger equation for a few spins exactly. We then describe in Sec.~\ref{sec:Rydberg_spins} how to implement the Heisenberg-XXZ spin model in a gas of ultracold atoms that are excited to Rydberg states. In  Sec.~\ref{sec:relaxation_dyn} we experimentally characterize the relaxation dynamics, and we theoretically investigate the dependence on disorder strength and character in Sec.~\ref{sec:hom_dyn}. Finally, we discuss in Sec.~\ref{sec:comparison} our findings in comparison to other systems exhibiting glassy dynamics.

\section{Qualitative picture of the quantum dynamics} 
\label{sec:qualitative_picture}

\begin{figure}[t!]
	\hspace*{-0.1cm}
	\includegraphics[width=0.9\columnwidth]{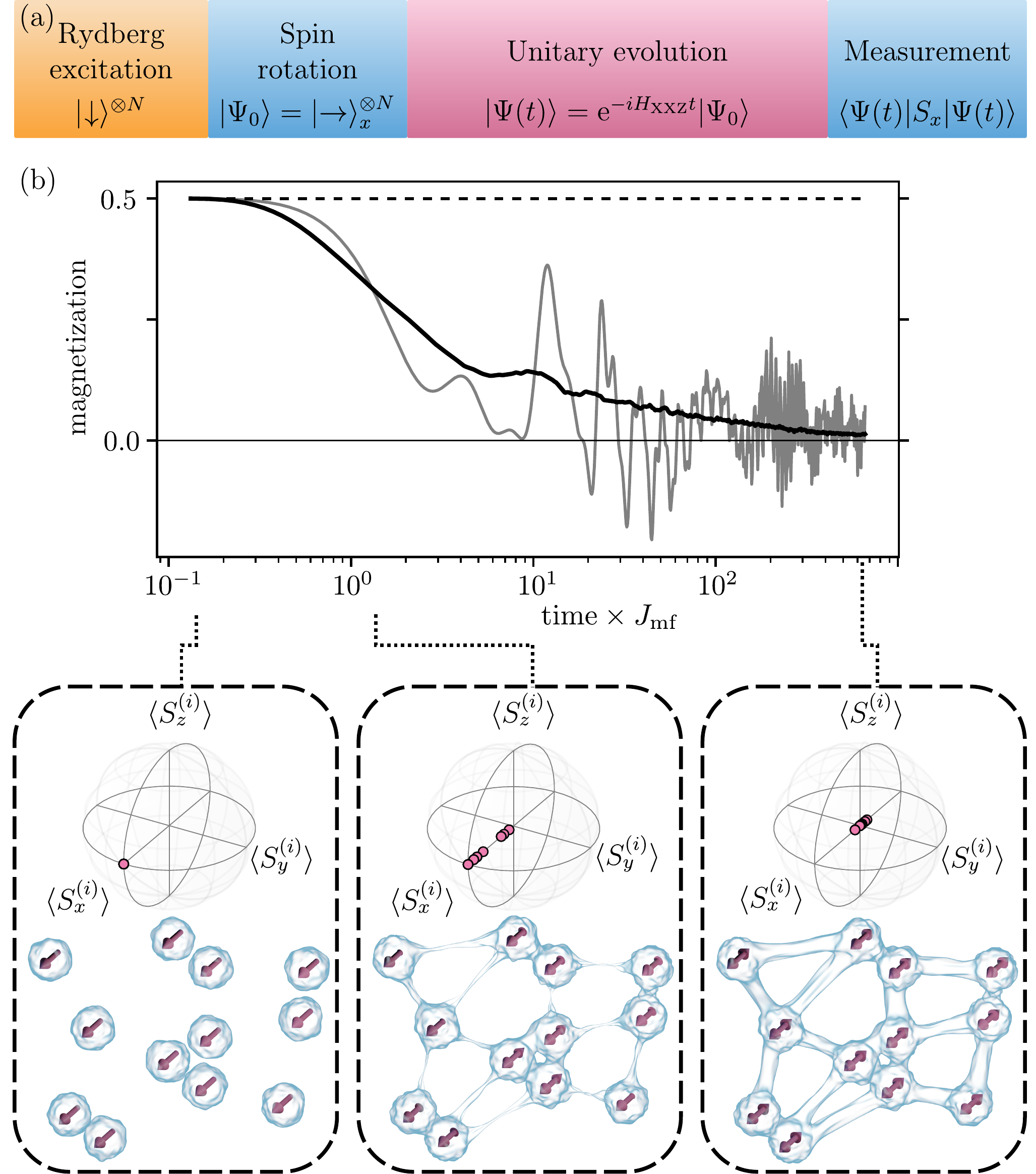}
	\caption{{Relaxation dynamics in a disordered quantum spin system.} (a) Protocol for initialization and readout of the many-body spin system composed of Rydberg atoms. Spin states $\ket{\uparrow}$ and $\ket{\downarrow}$ correspond to two different Rydberg states. (b) Exact simulation of 12 spins interacting via a Heisenberg XXZ Hamiltonian. The plot shows the magnetization for a single realization (gray curve) and the disorder average over 1000 realizations (black curve) which relaxes as function of time given in units of the median of the mean-field interaction strengths $J_{\mathrm{mf}}$. The dashed line indicates the mean-field prediction that does not relax. The microscopic expectation values of $\langle S_x^{(i)}\rangle$, $\langle S_y^{(i)}\rangle$ and $\langle S_z^{(i)}\rangle$ for each spin are plotted at three different time steps on the Bloch sphere. The reduction of the expectation values (magnetization) is a consequence of the spreading of entanglement (visualized by the blue bonds between spins). 
	}
	\label{fig:setup}
\end{figure}

We consider an ensemble of $N$ spin-1/2 particles randomly positioned in space and all initialized in the $\ket{\rightarrow}_x^{\otimes N} =1/\sqrt{2}(\ket{\uparrow} +\ket{\downarrow})^{\otimes N}$ state, corresponding to an initial magnetization $\langle S_x^{(i)}\rangle = 1/2$. Here, $S_{\alpha}^{(i)}$ ($\alpha = \{x, y, z\}$) refers to the spin-1/2 operator of the $i$th spin. The experimental protocol is illustrated in Figure~\ref{fig:setup}~(a). The unitary dynamical evolution of the system is governed by the Heisenberg XXZ-Hamiltonian in the absence of magnetic fields (in units where $\hbar=1$),
\begin{equation}
H_\mathrm{XXZ}=\frac{1}{2}\sum_{i,j} J_{ij}( S^{(i)}_x S^{(j)}_x+ S^{(i)}_y S^{(j)}_y + \delta S^{(i)}_z S^{(j)}_z)\,,
\label{eq:Hint}
\end{equation}
where $J_{ij}$ are the interaction couplings between the spins $i$ and $j$ and $\delta$ is the anisotropy parameter. To remain consistent with the experimental implementation (see Section~\ref{sec:Rydberg_spins}), we focus on an anisotropy parameter $\delta=-0.73$ and spin-spin interactions that decay as a power law $J_{ij} = {C_6}/{r_{ij}^6}$ with the inter-particle distance $r_{ij}$. 

To obtain a qualitative understanding of the quantum dynamics in this system, we perform a full quantum mechanical simulation on a small ensemble of $N=12$ spins. In Fig.~\ref{fig:setup}~(b) we show the time evolution of the magnetization $\langle S_x\rangle_k = 1/N \sum_i \langle S_x^{(i)} \rangle_k$ for a single disorder realization $k$ (grey curve). Due to the spatial disorder, spin-spin interactions give rise to complex many-body dynamics on strongly varying energy scales. 
This is in stark contrast to an effectively classical, mean-field prediction for this Hamiltonian [dashed line in Fig.~\ref{fig:setup}(c)], which assumes each spin to evolve in the average field generated by all other spins, thus neglecting quantum correlations. The initial fully magnetized state is an eigenstate of the mean-field Hamiltonian which explains the total absence of relaxation.

\begin{figure}[t!]
	\hspace*{-0.1cm}
	\includegraphics[width=0.65\columnwidth]{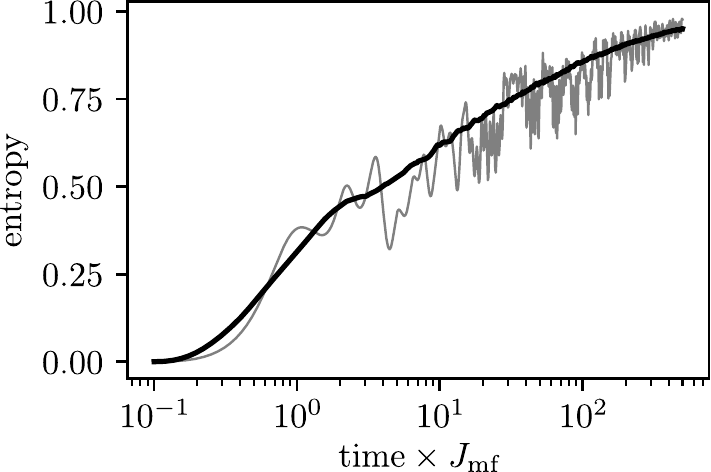}
	\caption{Build-up of entanglement quantified by the time evolution of the second order R\'{e}nyi entropy of the few-particle simulation from Fig.~\ref{fig:setup}(b). The plot shows the ensemble averaged entropy for a single realization (gray curve) and the disorder average over 1000 realizations (black curve). Since the full system remains pure, the R\'{e}nyi entropy is a measure of entanglement, that increases on similar timescales to the maximal value of $\mathcal{S}_i^{(2)}=1$ as the magnetization relaxes to zero.
	}
	\label{fig:purity}
\end{figure}

Therefore, in the many-body case the loss of magnetization is not caused by classical dephasing, but by the build-up of entanglement between spins, witnessed by the decrease of the local purity $\text{Tr}(\rho_i^2)$ for each spin $\rho_i$ being the single-spin reduced state $\rho_i=\text{Tr}_{\neg i} \rho$. Since the dynamics are unitary and therefore the full system remains pure ($\text{Tr}(\rho^2) = 1$), we can quantify entanglement by the second order R\'{e}nyi entropy
\begin{equation}
\mathcal{S}_i^{(2)} = -\log_2\left(\text{Tr}(\rho_i^2)\right),
\end{equation}
which increases to $\mathcal{S}_i^{(2)} = 1$ on a similar timescale as the relaxation of the magnetization (see Fig.~\ref{fig:purity}). After ensemble and disorder averaging, the magnetization approaches a fully randomized state with $\langle S_{x,y,z}\rangle=0$ [see black curve in Fig.~\ref{fig:setup}(b)], consistent with the ETH prediction. However, this relaxation occurs very slowly compared to the timescales associated with spin-spin interactions. 

\section{Realizing a Heisenberg-XXZ spin system with Rydberg atoms}
\label{sec:Rydberg_spins}

From a few-body perspective, one may wonder whether glassy dynamics can actually be observed in fully isolated quantum many-body systems and to what extent it shares common features with classical spin glasses. We address this question experimentally using a gas of ultracold rubidium atoms prepared in a superposition of two different Rydberg states. For well chosen pairs of states, the electric dipole-dipole coupling leads to the XXZ model~\cite{Whitlock2017,Nguyen2018}. Here we use the two low angular momentum states $\ket{\downarrow}=\ket{48s}$ and $\ket{\uparrow}=\ket{49s}$ to realize the Hamiltonian~\eqref{eq:Hint}, with $C_6/2\pi=\SI{59}{\giga\hertz\cdot\micro m \tothe{6}}$ characterizing the strength of the power law interactions (for the derivation of the Hamiltonian, see~\cref{sec:appxE_Rydberg interactions}). 

\begin{figure}[t!]
	\hspace*{-0.1cm}
	\includegraphics[width=1\columnwidth]{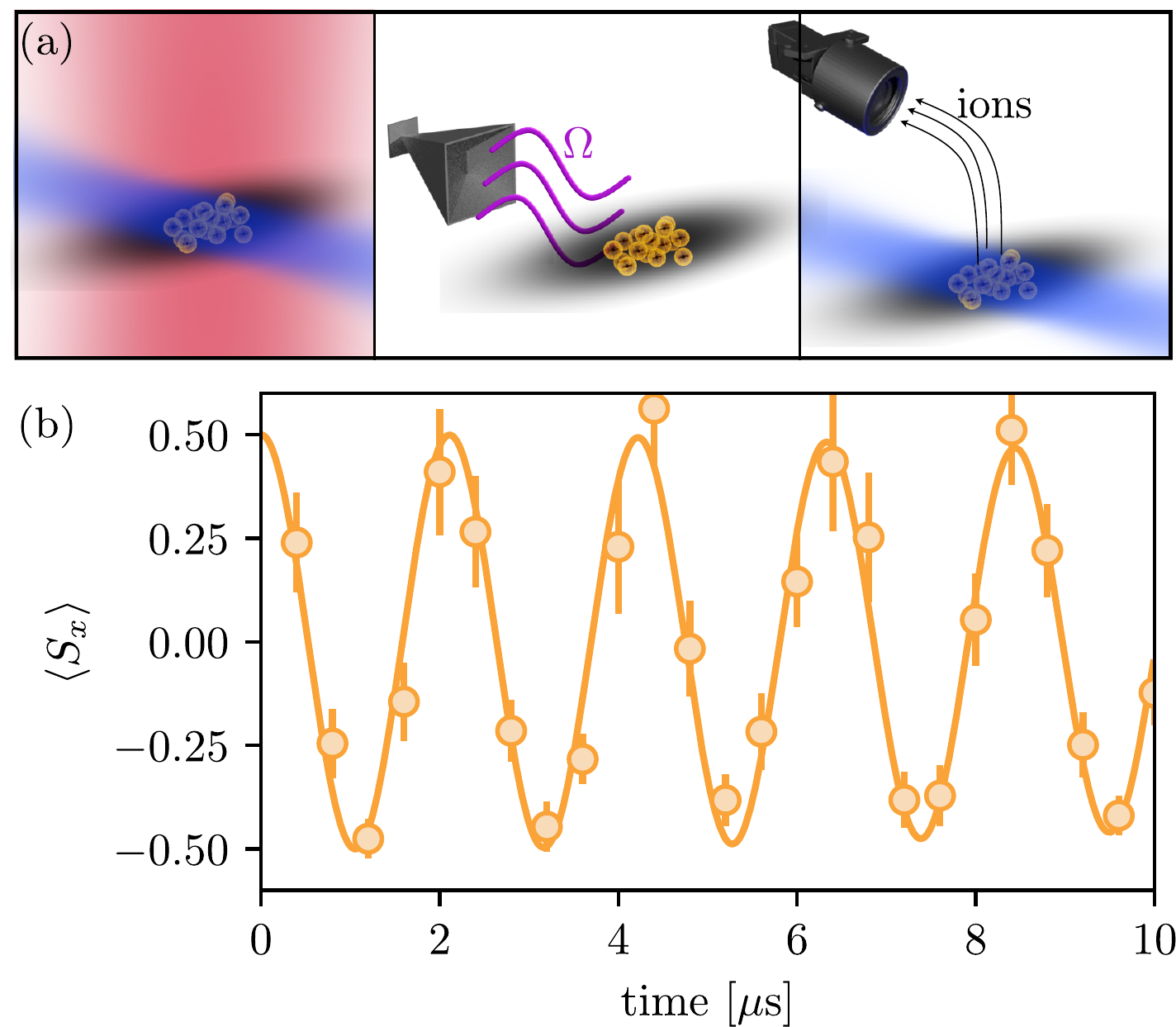}
	\caption{{Implementation of the XXZ spin-$1/2$ model in Rydberg gases.}
		(a) Experimental procedure. Left: A laser pulse at \SI{780}{\nano\meter} (red) and \SI{480}{\nano\meter} (blue) excites a controlled fraction of the $^{87}\text{Rb}$ atoms to the $\ket{\downarrow} = \ket{48S}$ Rydberg state. Middle: A microwave field couples the $\ket{\downarrow}$ state to the $\ket{\uparrow} = \ket{49S}$ state to perform a Ramsey experiment. Right: A blue \SI{480}{\nano\meter} laser depopulates the $\ket{\downarrow}$ Rydberg state, before the $\ket{\uparrow}$ state is ionized by an electric field. The ions are detected by a Multichannel Plate (MCP). (b) Ramsey fringes showing a high degree of phase coherence for a detuning of $\Delta/2\pi=\SI{0.47}{\mega\hertz}$ for a peak spin density of $\rho_{\mathrm{S}}^0 = 6.0(15) \times 10^7 \text{cm}^{-3}$. The solid line shows discrete Truncated Wigner Approximation simulations for that density.}
	\label{fig:time_ramsey}
\end{figure}

The experimental procedure [Fig.~\ref{fig:time_ramsey}~(a)] starts with a gas of $^{87}$Rb atoms prepared in their electronic ground state $\ket{g}=\ket{5S_{1/2},F=2,m_F=2}$ in an optical dipole trap and with a temperature T $\sim 50\,\mu$K, low-enough to freeze the motional degrees of freedom over the time scale of the experiment. A laser pulse of variable duration brings a controllable number of atoms $N\leq 1200$ to the $\ket{\downarrow}= \ket{48S_{1/2},m_j=+1/2}$ Rydberg state. For this we use a two-photon laser excitation at $\SI{780}{\nano\meter}$ and $\SI{480}{\nano\meter}$, with a detuning $-2\pi\times \SI{100}{\mega\hertz}$ from the intermediate state $\ket{e}=\ket{5P_{3/2},F=3,m_F=3}$ and an effective Rabi frequency of $2\pi\times \SI{150}{\kilo\hertz}$. To individually address two specific Rydberg states, including Zeeman substructure, we continuously apply a magnetic field of \SI{6}{\gauss}. To characterize the resulting three-dimensional Gaussian Rydberg density distribution we perform depletion imaging, where the Rydberg density is deduced by absorption imaging of the ground-state atoms before and after the Rydberg excitation laser pulse~\cite{ferreira2020depletion}.

A two-photon microwave field is then used to couple the $\ket{\downarrow}$ state to the $\ket{\uparrow}=\ket{49S_{1/2},m_j=+1/2}$ Rydberg state. The single photon frequency $\nu=\SI{35.2}{\giga\hertz}$ is detuned from the intermediate state $\ket{48P_{3/2}}$ by $\SI{170}{\mega\hertz}$, far enough to guarantee that the population in this state due to off-resonant coupling is smaller than 2\%. The atoms can therefore be considered as two-level systems described by a pseudo-spin degree of freedom. In this description the microwave field acts as an external field described by the Hamiltonian
\begin{equation}
H_\mathrm{ext} = \sum_i \left(\Omega \sin{\phi} S^{(i)}_x -\Omega \cos{\phi} S^{(i)}_y + \Delta S^{(i)}_z\right)\,,
\label{eq:H}
\end{equation}
with $\Omega$ the Rabi frequency, $\Delta$ the two-photon detuning and $\phi$ the phase of the field. The Rabi frequency is calibrated from the period of Rabi oscillations between the two spin states.

To perform the Ramsey sequence shown in Fig.~\ref{fig:setup}~(a), a resonant microwave $\pi/2$ pulse at an effective Rabi frequency $\Omega/2\pi = \SI{3.00 \pm 0.01}{\mega\hertz}$ rotates all spins to the fully-magnetized state $\ket{\rightarrow}_x^{\otimes N}$ with all spins pointing along the $x$-direction on the Bloch sphere. Uncertainties in the duration and amplitude of the pulses as well as interaction effects lead to imperfect initial spin state preparation. Based on simulations, we estimate the fidelity to be higher than 96\%. 

After a free evolution time $t$ in the absence of the microwave field ($\Omega=0$), a second microwave $\pi/2$ pulse with adjustable phase $\phi$ is applied to rotate the equatorial magnetization components
\begin{equation}
\langle S_\phi \rangle = \cos{\phi} \langle S_x \rangle + \sin{\phi} \langle S_y \rangle
\label{eq:S_phi}
\end{equation}
to the detection basis $\{\ket{\downarrow},\ket{\uparrow}\}$. In this way we effectively read out the $\langle S_x \rangle$ and $\langle S_y \rangle$ magnetizations from population measurements of the Rydberg states using electric field ionization (see \cref{sec:Appx_Determination_Magnetization}).

To ensure unitary Hamiltonian dynamics, we restrict the experimental time scales to a maximum of \SI{10}{\micro\second} which is short compared to the spontaneous decay time and redistribution by black-body radiation (\SI{113}{\micro\second} and \SI{121}{\micro\second} respectively for the chosen Rydberg states~\cite{Sibalic2017}).  To verify that the single-spin phase coherence is preserved during experimental time, we perform a Ramsey measurement with finite detuning $\Delta$ at low spin densities where interactions can be neglected~\cite{Carter2013,Hermann2014}. The full contrast oscillation shows that the single-spin phase coherence is preserved over the duration of the experiment, as shown in Fig.~\ref{fig:time_ramsey}(b).

\section{Experimental observation of relaxation dynamics}
\label{sec:relaxation_dyn}

\subsection{Glassy dynamics}

\begin{figure}[tbp!]
	\hspace*{-0.1cm}
	\includegraphics[width=0.9\columnwidth]{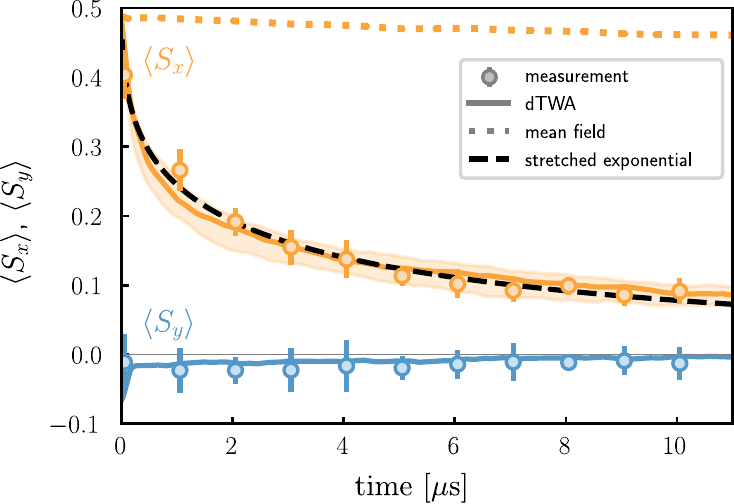}
	\caption{{Many-body relaxation dynamics of a Heisenberg XXZ Rydberg spin system.} 
		The dots show the temporal evolution of the two magnetization components $\langle S_x \rangle$ and $\langle S_y \rangle$ from a tomographic spin readout. Error bars are determined from 120 realizations of the experiment at a peak spin density of $\rho_{\mathrm{S}}^0 = \SI{1.2(3)e9}{\centi\meter \tothe{-3}}$. The observed dynamics are clearly inconsistent with the mean-field prediction including imperfect initial state preparation (dotted line). The solid lines are dTWA predictions without free parameters. The shaded area indicates the systematic uncertainty of the measured density. The dashed line depicts a fit of the data with a stretched exponential function yielding a stretching exponent of $\beta= 0.32(2) $.
	}
	\label{fig:time}
\end{figure}

We now study the relaxation dynamics due to spin-spin interactions for increasing spin densities. Fig.~\ref{fig:time} shows the experimentally observed relaxation of the magnetization using tomographic spin-resolved readout of $\langle S_x\rangle$ and $\langle S_y\rangle$. Starting from the almost fully magnetized state $\langle S_x\rangle=1/2,\,\langle S_y\rangle\approx0$ we observe that the magnetization decays towards the unmagnetized state $\langle S_x\rangle=0,\,\langle S_y\rangle=0$ within $\approx \SI{10}{\micro\second}$. This is much shorter than the single-spin phase coherence time measured in Fig.~\ref{fig:time_ramsey} but still slower than the characteristic timescale of interactions, $\left(C_6/(2\pi)/a_0^6\right)^{-1}=\SI{0.7(3)}{\micro\second}$. Here, $a_0 = \left(4\pi\rho_\text{S}^0/3\right)^{-1/3}$ represents the mean inter-particle distance. It is defined as the Wigner-Seitz radius of the Gaussian spin density distribution~\cite{Hertz1909}, whose peak density $\rho_S^0$ is given by the initial peak density of spin-down Rydberg atoms.

The time evolution seen in the experiment is qualitatively similar to the dynamics obtained by exact diagonalisation in Fig.~\ref{fig:setup}(b). Even when accounting for imperfect preparation of the initial state $\langle S_x\rangle\lesssim 1/2$ the mean-field prediction shows essentially no relaxation on the experimentally relevant timescales (see dotted line in Fig.~\ref{fig:time} and~\cref{sec:appxF_theoretical_models}). 
The qualitative failure of the mean-field description is different from our earlier observation~\cite{Orioli2018} of a density-dependent dephasing of $\langle S_z \rangle$. In this previous work, mean field also predicted a damping of Rabi oscillations, yet failing to provide a quantitatively consistent description of the dynamics. Here, the absence of dynamics at the mean-field level implies that the relaxation seen in the experiment is closely related to build-up of entanglement also apparent in the exact diagonalisation calculations. 

We find that the relaxation is well described by a stretched exponential, in apparent similarity with glassy dynamics in classical disordered media~\cite{Phillips1996, Choi2017, Fischer2016, Mukherjee2016}: 
\begin{equation}
\langle S_x(t) \rangle = \frac{1}{2} \exp[-(\gamma_J t)^\beta]\,
\label{eq:stretched}
\end{equation}
where $\beta$ is the stretching exponent and $\gamma_J$ defines an effective relaxation rate. The exponent $\beta$ characterizes the deviation from a simple exponential ($\beta=1$) towards a purely logarithmic decay ($\beta\rightarrow 0$) [see~\cref{sec:appxG_Stretched_Exponential}].
The experimental data are well described by this phenomenological function (dashed line in Fig.~\ref{fig:time}) yielding an exponent $\beta=0.32(2)$. This value clearly rules out a pure exponential decay, i.e. $\beta=1$, that could be expected on the basis of single-particle dephasing.

\subsection{Insensitivity to microscopic details}
\label{sec:universal_dynamics}

\begin{figure}[t!]
	\hspace*{-1.5cm}
	\includegraphics[width=0.8\columnwidth]{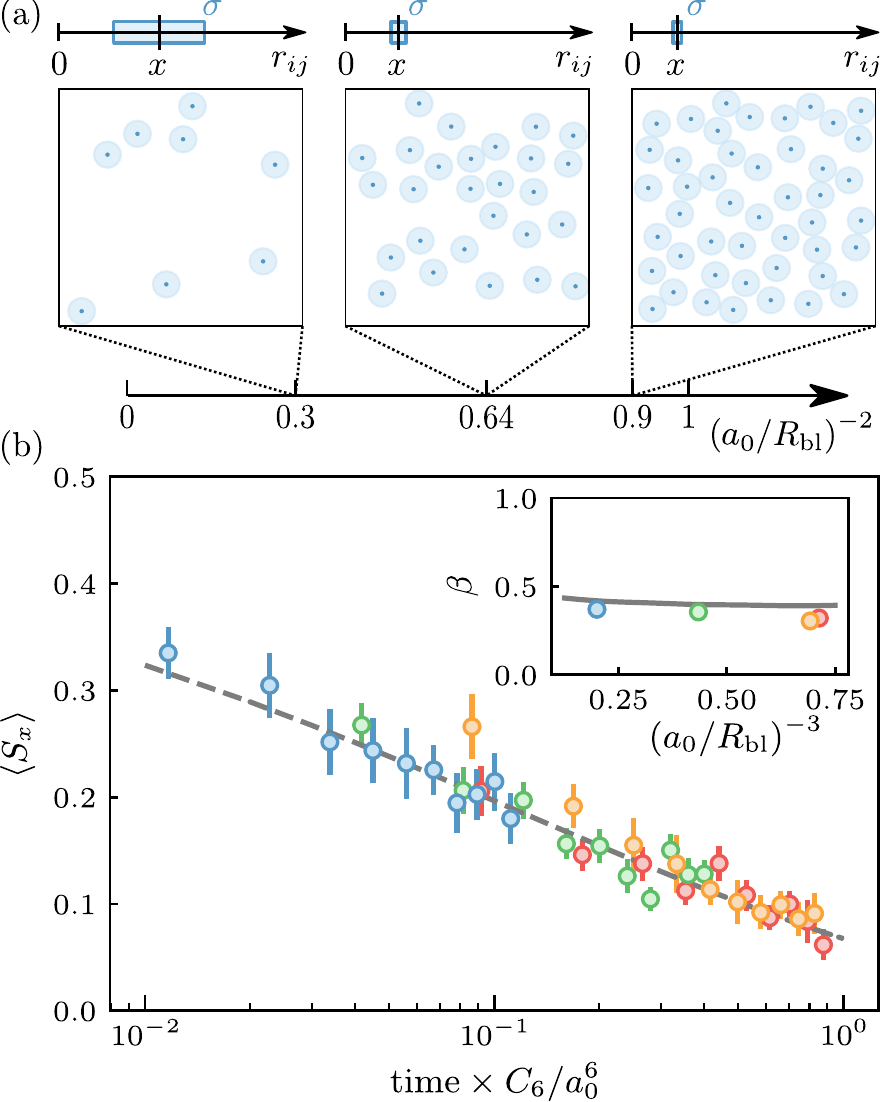}
	\centering
	\caption{{Rescaled magnetization dynamics for different densities and disorder strengths.} (a) 2D representation of a spin system with different densities and thus strengths of disorder, characterized by the mean number of spins per blockade radius $\left({a_0}/{R_{\text{bl}}}\right)^{-2}$. The bar denotes the mean $x$ and standard deviation $\sigma$ of the interparticle-distances $\min_{i\neq j} r_{ij}$. (b) Data points represent measurements of the averaged magnetization $\langle S_x \rangle $ for spin densities $\rho_{\mathrm{S}}^0= 0.43(11) \times 10^9 \text{cm}^{-3}$ (blue), $ 0.8 (2) \times 10^9 \text{cm}^{-3}$ (green),  $ 1.2 (3) \times 10^9 \text{cm}^{-3}$ (orange and red, corresponding to two different data sets). By rescaling time with the effective interaction strength $C_6/a_0^6$ the data collapses on a single curve described by a stretched exponential function with stretching exponent  $\beta = 0.32(2)$ (dashed line). The inset shows $\beta$ as a function of the corresponding ratio $\left({a_0}/{R_{\text{bl}}}\right)^{-3}$ for the experimental data and dTWA calculations (solid gray line).
	}
	\label{fig:inhom-rel}
\end{figure}

To further investigate how slow relaxation and the characteristic exponent depends on microscopic details, we control the degree of spatial disorder by taking advantage of the Rydberg blockade effect in the state preparation stage~\cite{Schauss2012,Bettelli2013,Urvoy2015}. During laser excitation the strong van der Waals interactions between Rydberg states prevents two spins from being prepared at distances smaller than the Rydberg blockade radius $R_{\text{bl}}$. The degree of disorder is thus controlled by the ratio between blockade radius $R_{\text{bl}}$ and Wigner-Seitz radius $a_0$ (Fig.~\ref{fig:inhom-rel}a). For $a_0 \gg R_{\text{bl}}$ the blockade effect has little influence and the spins are randomly distributed, whereas the limit $a_0\approx R_{\text{bl}}$ corresponds to a strongly ordered configuration. In between, the short distance cutoff imposed by the Rydberg blockade effect effectively reduces the strength of the disorder compared to fully uncorrelated random spin positions.

In the experiment we can tune the disorder strength by changing the peak spin density $\rho_{\mathrm{S}}^0$ and thus the mean number of spins per blockade sphere $(a_0/R_\text{bl})^{-3}$ from $0.20(5)$ to $0.7(2)$ (two-dimensional representations of corresponding distributions are depicted in Fig.~\ref{fig:inhom-rel}~(a)). 
Remarkably, we find the stretching exponent $\beta$ to be almost constant over this range (inset in Figure \ref{fig:inhom-rel}b). Furthermore, after rescaling the time axis by the characteristic energy scale $C_6/a_0^6$, the time-dependent data collapse onto a single line (Fig. \ref{fig:inhom-rel}~(b)). From this we conclude that the dynamics is insensitive to the disorder strength which is modified by the blockade effect. These experimental observations are indicative of a universal behaviour in the sense that the dynamics does not depend on the microscopic details of the system. This unexpected feature will be explored further in numerical simulations.

Due to the large number of spins in the experiment, an exact computer simulation of the unitary dynamics under the Hamiltonian Eq. (\ref{eq:Hint}) is not possible. Instead, quantum effects can be partially taken into account by applying the semiclassical discrete Truncated Wigner Approximation (dTWA, see~\cref{sec:appxF_theoretical_models})~\cite{Polkovnikov2010,Schachenmayer2015} which has recently been shown to describe the dynamics of Rydberg interacting spin systems very well~\cite{Orioli2018}. To model the present experiments, all physical parameters entering the simulation are determined through independent measurements, such as the spatial density distribution, total number of spins and the microwave coupling strength $\Omega$ used in the preparation and readout stages. The initial spin distribution is generated from a random excitation model of the Rydberg atoms, including a cut-off distance to account for the blockade effect. This classical sampling of the spatial spin distribution is justified on the basis that neither the microwave pulses nor the Rydberg-Rydberg interactions couple different terms of the collectively excited many-body state, each of which satisfy the blockade constraint (see~\cref{sec:appxD_spatial_distribution}). The numerical simulations describe the glassy dynamics and the insensitivity with respect to changes of the distribution function of the interaction strength very well [solid line in Fig.~\ref{fig:time} and in the inset of Fig.~\ref{fig:inhom-rel}(b)], further confirming the validity of the dTWA approximation for treating the dynamics of disordered quantum systems.

\section{Numerical study of the role of disorder strength}
\label{sec:hom_dyn}

\begin{figure}[t!]
	\hspace*{-0.1cm}
	\includegraphics[width=1\columnwidth]{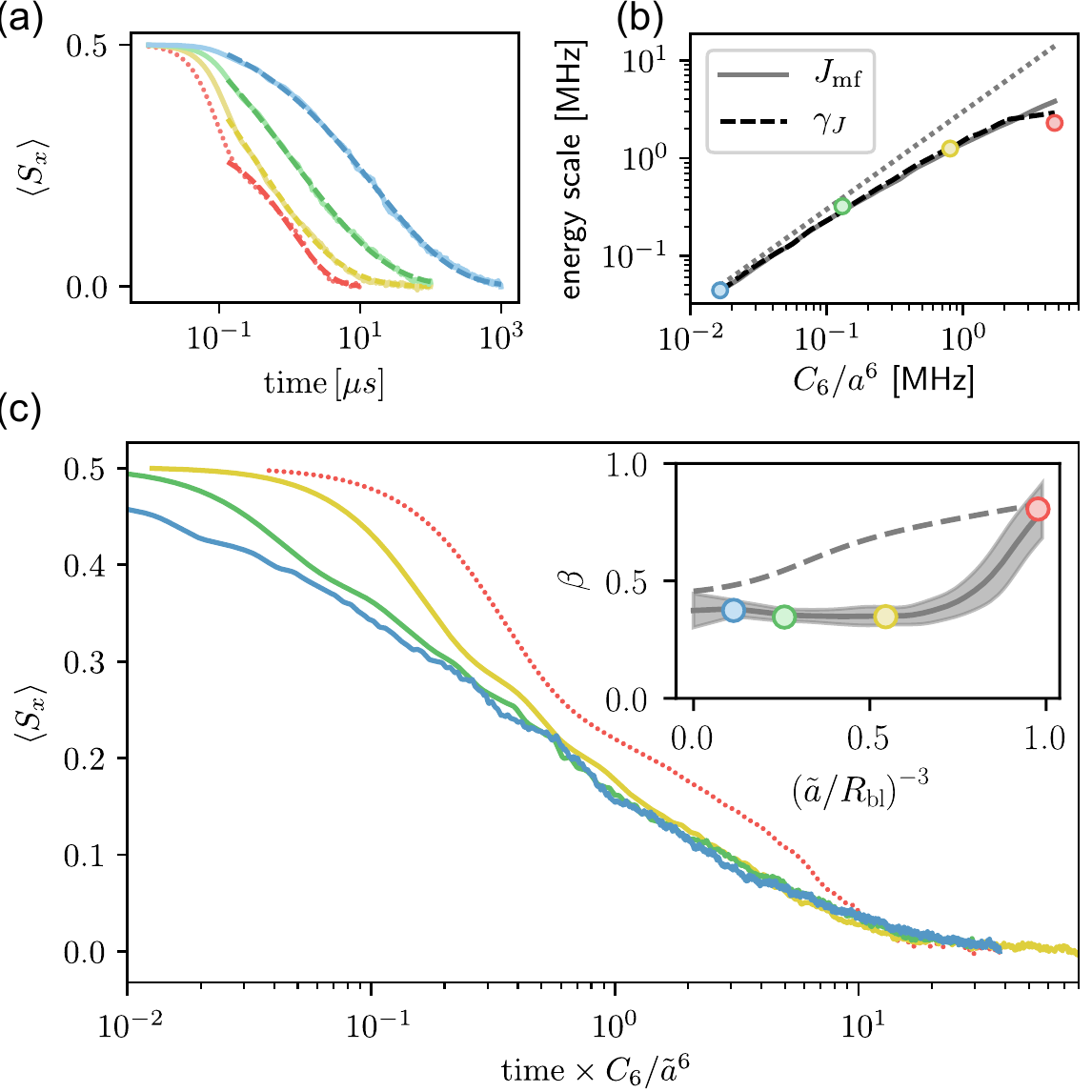}
	\caption{Numerical simulation of the dynamics for a uniform density distribution using the dTWA. 
		(a) The simulated dynamics before rescaling. After the quadratic onset, the magnetization decay is fitted well by the stretched exponential function (dashed lines). (b) The fitted decay rate $\gamma_J$ (dashed black line) agrees well with the median of the mean-field interaction strengths $J_{\text{mf}}$ (solid grey line), which is the typical energy scale of the system. In the weakly interacting limit, $J_{\text{mf}}$ scales linearly with $C_6/a^6$ (dotted grey line). The colored dots denote the decay rate derived from the four simulations depicted in (a). (c) For small densities of $\rho_{\mathrm{S}}= 1.25 \times 10^8 \text{cm}^{-3}$ (blue line and blue dot in the inset), intermediate densities of $ 3.51 \times 10^8 \text{cm}^{-3}$ (green line and green dot in inset) and $ 8.73 \times 10^8 \text{cm}^{-3}$ (yellow line and yellow dot in the inset), the numerical data collapses on one curve after rescaling time with $C_6/\tilde{a}^6$, where $\tilde{a}$ plays the role of an effective distance that takes into account the roles of disorder and power law interactions. For densities as large as $\rho_{\mathrm{S}}= 2.11 \times 10^9 \text{cm}^{-3}$ (dotted red line and red dot in the inset), the spatial order introduced by the blockade is so large that the dynamics does not follow the universal behavior observed for smaller densities. 
		Inset: Disorder averaged stretching exponent $\beta$ (solid line) as a function of the order in the system expressed by $\left(\tilde{a}/R_{\text{bl}}\right)^{-3}$. Below a critical value of $\left(\tilde{a}/R_{\text{bl}}\right)^{-3} \lesssim 0.7$ the exponent becomes constant. The shaded area indicates statistical uncertainty. The dots denote the exponent $\beta$ resulting from the single disorder realizations of panel (a). The dashed line is obtained from the fluctuator model (see eq.~\ref{eq:rate_equation} in section~\ref{sec:comparison}).
	}
	\label{fig:hom-rel}
\end{figure}

Theoretical modeling using the dTWA allows to further test the role of disorder for the glassy dynamics, while excluding possible effects of the inhomogeneous spin density resulting from the optical trap. The simulated spin dynamics for a uniform density distribution $\rho_{\text{S}}$ are shown by the solid lines in Fig.~\ref{fig:hom-rel}(a). For early times where $t \le 2\pi R_{\text{bl}}^6/C_6$, the leading order quadratic Hamiltonian evolution is clearly visible. For times beyond the perturbative short-time regime, the relaxation is well described by a stretched exponential [dashed lines in Fig.~\ref{fig:hom-rel}(a), see~\cref{sec:appxG_Stretched_Exponential}], proving that the glassy dynamics is an intrinsic many-body effect and not a result of the density inhomogeneities.

Fig~\ref{fig:hom-rel}(b) shows the fitted relaxation rate $\gamma_J$ of the four simulations from Fig~\ref{fig:hom-rel}~(a) (colored dots, the disorder average is shown as a dashed black line). This rate does not scale with $C_6/a^6$ [$a = (4\pi \rho_{\text{S}}/3)^{-1/3}$ (dotted line)] but with the median of the mean-field interaction strengths $J_{\text{mf}}$ (solid line), which plays the role of the characteristic energy scale in the system (see~\cref{sec:appxD_spatial_distribution}). This scale $J_{\text{mf}} = C_6/\tilde{a}^6$ defines an effective Wigner-Seitz radius $\tilde{a}$, that coincides with the usual Wigner-Seitz radius $a$ for small spin densities but deviates at larger densities when spatial correlations induced by the Rydberg blockade effect become important. Rescaling the time with $C_6/\tilde{a}^6$, the simulated data from Fig.~\ref{fig:hom-rel}(a) collapse on a single stretched exponential curve [see Fig.~\ref{fig:hom-rel}(c)], similarly to the experimental observations in Fig.~\ref{fig:inhom-rel}(b). 

This insensitivity to changes in the cut-off energy induced by the blockade effect substantiates the nature of the universal relaxation dynamics. This is further confirmed by the fitted stretching exponent $\beta$ shown in the inset of Fig.~\ref{fig:hom-rel}(c) obtained after disorder averaging [the solid line depicts the mean, the grey shaded area shows the standard deviation, the colored dots the single disorder realizations from Fig.~\ref{fig:hom-rel}(a)]. We find $\beta$ to be approximately constant for large disorder strengths where spatial correlations are weak (i.e. $\left(\tilde{a}/R_{\text{bl}}\right)^{-3} \lesssim 0.7$), with a value of $0.36$, close to the experimental value of $\beta=0.32(2)$. In fact, the range of densities in the inhomogeneous experimental distribution falls well into the regime of a constant stretching exponent (see inset in Fig.~\ref{fig:inhom-rel}~(b)). 

\begin{figure}[thb]
	\hspace*{-0.1cm}
	\includegraphics[width=0.9\columnwidth]{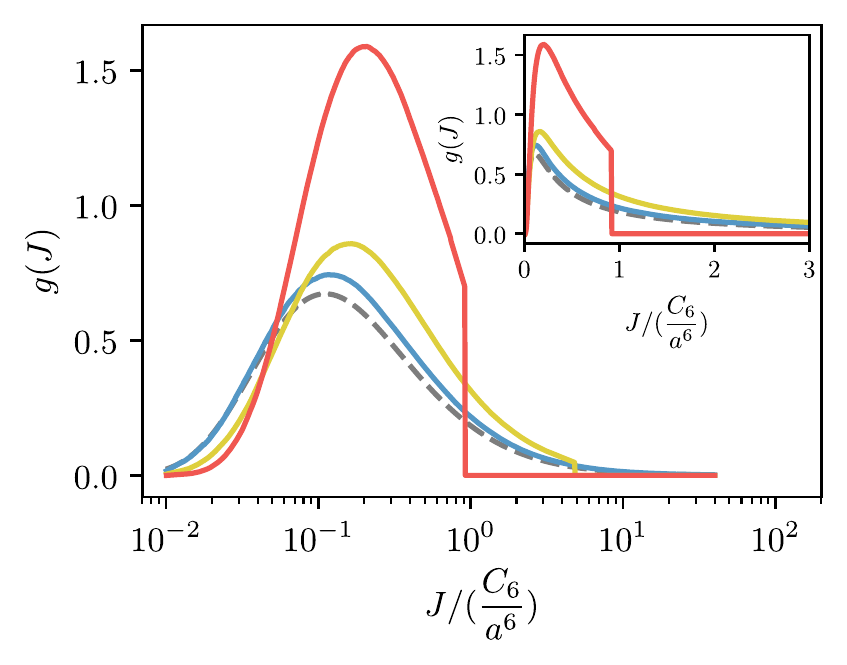}
	\caption{Normalized probability distribution of nearest-neighbour interactions $g(J)$ for the densities given in Fig.~\ref{fig:hom-rel} (blue line: $\rho_{\mathrm{S}}=1.25 \times 10^8 \text{cm}^{-3}$, yellow line: $\rho_{\mathrm{S}} = 8.73 \times 10^8 \text{cm}^{-3}$, red line: $\rho_{\mathrm{S}}= 2.11 \times 10^9 \text{cm}^{-3}$). The dashed line depicts a distribution corresponding to randomly placed spins. The blockade effect induces correlations in the system and hence the probability distribution at high densities become more peaked. This results in a decreased disorder compared to the completely random case. The inset shows the same data with linear axis to illustrate the narrowing of the distribution functions. 
	}
	\label{fig:DKL}
\end{figure}

The simulations allow to access even higher densities than those accessible in the experiment, corresponding to more correlated spatial configurations of the atoms. For $(\tilde{a}/R_{\mathrm{bl}})^{-3}=1$, the dTWA simulations show significantly different dynamics [dotted line in Fig.~\ref{fig:hom-rel}(a)] that translates into a stretching exponent $\beta$ that becomes sensitive to the strength of the disorder above a certain threshold [see inset in Fig.~\ref{fig:hom-rel}(c)]. 

To understand the behaviour of $\beta$ when changing the disorder, we analyse the influence of the blockade effect on the distribution of coupling strengths. Figure~\ref{fig:DKL} shows for four different densities the distribution function of nearest-neighbor interaction strength rescaled by $C_6/a^6$. At low densities, where we observe a disorder-independent stretching exponent, the rescaled distributions do not fully coincide due to the blockade effect but their shape remains qualitatively the same. Their functional form is mostly modified at large interaction strengths which influences only the short time dynamics. At large density, however, the distribution is strongly altered by the high-energy cut-off which renders the distribution much more peaked.

It is thus tempting to conjecture that the insensitivity of the dynamics below a certain disorder strength is related to negligible changes in the distribution of interaction strength. Nonetheless, the changes at low density are significant enough to predict a non-constant $\beta$ when applying a simple model such as the fluctuator model discussed in the next section, ruling out such a simple interpretation based on energy rescaling.

\section{Comparison to other systems exhibiting glassy dynamics}
\label{sec:comparison}

It is instructive to compare our findings to other systems exhibiting glassy dynamics. In \emph{isolated} quantum systems, glassy dynamics has been observed for the Ising model which is exactly solvable by the Emch-Radin ansatz~\cite{Emch1966,Radin1970,Mukherjee2016}. In this analytical solution the averaging over oscillatory terms results in a stretching exponent of $\beta=1/2$ for disordered spin systems which was also confirmed experimentally~\cite{Takei2016, Sommer2016}. 
Such a treatment is not possible for an arbitrary Heisenberg XXZ-Hamiltonian since the Emch-Radin ansatz is based on the commutativity between the terms in the Hamiltonian, which is only given for the Ising model. Since the additional exchange interactions of the Heisenberg XXZ model make the system inaccessible to analytical treatment, quantum simulations are essential to investigate the problem for a large number of spins~\cite{Hazzard2013}. In this respect, our observation of slow dynamics in a Heisenberg XXZ-model in 3D with long-range interactions reveals that glassy dynamics is not an extraordinary property of the special case of Ising systems, but a more generic feature applicable to isolated disordered spin systems.
In our quantum simulation experiment, we observed a value of $\beta=0.32(2)$, significantly lower than $1/2$.

A more general approach to explaining glassy dynamics can be deduced from \emph{open} disordered systems where sub-exponential relaxation is a ubiquitous phenomenon~\cite{Phillips1996}. Here, the incoherent averaging over a distribution of decay rates leads to the stretched exponential. The origin giving rise to a specific distribution depends on the exact physical system or model.

A commonly used model that predicts glassy dynamics is the fluctuator model, where each spin is embedded in a local bath that determines its decay rate. For example, this model has been applied to NV centers where spins couple to randomly distributed fluctuators~\cite{Choi2017,Choi2017_tc, Kucsko2018}. A similar approach is taken for dissipative many-body localized systems where random decay rates originate from the disordered potential at each lattice site~\cite{Everest2017, Luschen2017}. For our system, we apply the fluctuator model assuming that the relaxation of each spin can be effectively described by an incoherent coupling to its environment, which is determined by the local rate $\gamma_i$ sampled from the probability distribution of nearest neighbor interactions from Fig.~\ref{fig:DKL}\footnote{Qualitatively the same results are obtained for sampling the decay rates from the distribution of mean-field interaction strengths}:
\begin{equation}
\langle S_x^{(i)}(t)\rangle = \frac{1}{2}\exp\left[ - \gamma_i t \right]
\label{eq:rate_equation}
\end{equation}
This model predicts a stretched exponential relaxation as observed in the experiment. However, we find two discrepancies: First, the model predicts stretching exponents of $\beta > 1/2$ inconsistent with our previous finding of $\beta$ being smaller than $1/2$. Second, the fluctuator model predicts a stretching exponent varying as a function of disorder strength, as shown by the dashed line in the inset of Fig.~\ref{fig:hom-rel}~(c). 
Hence this model is sensitive to the change in the distribution functions of the interaction strength introduced by the cut-off as depicted in Fig.~\ref{fig:DKL}, already in the regime of large disorder. Therefore, the fluctuator model does not capture the collapse of dynamics onto a single curve after rescaling.

The most prominent example of open systems exhibiting slow dynamics are spin glasses where, at low temperatures, the relaxation of the macroscopic magnetization is described by a temperature dependent stretching exponent~\cite{Binder1986, Bouchaud1998}. Similar to the fluctuator model, this can be explained by incoherent averaging over random relaxation times. However, these times do not result from a local environment of each spin, but from a random distribution of free energies~\cite{DeDominicis1985}. In this approach, concepts from equilibrium statistical mechanics like thermodynamic potentials are applied to describe the long-time evolution of the system.
Although our system does not exhibit sign-changing interactions~\cite{young1996numerical} or contains geometric frustration~\cite{Crespo2013} characteristic for spin glasses,
the observation of stretched exponential decay indicates that at long times a similar quasi-equilibrium approach might be applicable. This conjecture is supported by the fact that our system is expected to thermalize in the sense of the ETH, but approaches this limit only at exponentially long times in analogy to classical spin glasses.

\section{Conclusion}

In this work we implemented the XXZ Hamiltonian in 3D using a frozen gas of Rydberg atoms. We studied the out-of-equilibrium dynamics of this model starting from an almost zero-entropy initial state, which lacks a thorough theoretical understanding and is hard to simulate by classical means. We observed glassy dynamics in close analogy to the subexponential relaxation known from open disordered systems described by a stretched exponential function. While the latter is driven by thermal fluctuations, the dynamics of the disordered isolated quantum system is governed by quantum fluctuations and spreading of entanglement going beyond mean-field approximations. The observation that the dynamics of the magnetization is well described by semi-classical truncated Wigner simulations suggests that quantum interference effects become less important as the system approaches its equilibrium state. This is in line with previous findings that the long-time dynamics of generic thermalizing quantum many-body systems simplifies also in the sense that states can be represented efficiently due to limited entanglement~\cite{Leviatan2017}.

In the experiment, disorder is changed by exploiting the Rydberg blockade, which shifts the upper cutoff scale in the distribution of interaction strengths. Remarkably, the stretching exponent $\beta$ takes on a constant value above a certain disorder strength, as confirmed by both experiment and semi-classical simulations.  The long time evolution is therefore insensitive to the microscopic details of the system parameters on high energy scales. 
We interpret the independence of the dynamics to changes in the distribution function of interaction strengths as universal behaviour.  This and the validity a of semi-classical description as a strong hint that the dynamics of many-body quantum systems might be amenable to a simplified description of the late-time dynamics in terms of effective low energy degrees of freedom. Concretely, this could be approached within the framework of the strong disorder renormalization group, iteratively integrating out the highest energy degrees of freedom resulting from most strongly interacting spins or clusters of spins~\cite{Igloi2018}. Furthermore, spin glasses in the aging regime have been found to show certain quasi-thermal properties. For examples a fluctuation dissipation theorem has been found to hold~\cite{Herisson2002} and the spin-glass transition shows similarities to a thermal phase transition~\cite{Anderson1988}. Thus, the similarities to dynamics in open glassy systems observed in this work are encouraging to extend such an effective thermal-like description to \emph{quantum} glassy dynamics.

\section*{Acknowledgements} \label{sec:acknowledgements}

We thank J. Berges, A. Pi\~{n}iero Orioli, M. Rigol and J. Schachenmayer for fruitful discussions. This work is part of and supported by the DFG Collaborative Research Centre "SFB 1225 (ISOQUANT)", the DFG Priority Program "GiRyd 1929", the European Union H2020 projects FET Proactive project RySQ (Grant No. 640378) and FET flagship project PASQuanS (Grant No. 817482) and the Heidelberg Center for Quantum Dynamics. S.W. acknowledges support by ``Investissements d'Avenir'' programme through the Excellence Initiative of the University of Strasbourg (IdEx). T.F. acknowledges funding by a graduate scholarship of the Heidelberg University (LGFG) and  R.F. from the Brazilian fund Ci\^{e}ncia sem Fronteiras.

\appendix

\section{Calculation of Rydberg interactions and spin model}
\label{sec:appxE_Rydberg interactions}

In order to describe the interaction between two Rydberg excitations, the Hamiltonian is expanded in multipoles. This is well justified, as the minimal distance between the Rydberg atoms that is determined by the blockade radius $R_{\text{bl}}$ is much larger than the LeRoy radius $R_{\text{LR}}$ that describes the typical spread of the electron wave function. The leading order term of this expansion is the dipole-dipole interaction Hamiltonian
\begin{equation}
\hat{H}_{\text{DDI}} = \frac{\boldsymbol{\hat{d}}_i \cdot \boldsymbol{\hat{d}}_j - 3\left( \boldsymbol{\hat{d}}_i \cdot \boldsymbol{e}_r\right)\left( \boldsymbol{\hat{d}}_j \cdot \boldsymbol{e}_r \right)}{R^3}
\label{eq:dipoleground}
\end{equation}
that couples Rydberg atoms with different angular moment quantum number $l$. For dipolar forbidden transitions, the second order term in perturbation theory needs to be calculated giving rise to the van der Waals Hamiltonian
\begin{equation}
H_{\text{vdW}} = -\frac{1}{\hbar} 	\sum_m \frac{H_{\text{DDI}}|m\rangle\langle m|H_{\text{DDI}}}{\Delta_F} \delta(\omega_{fm} + \omega_{mi})
\label{eq::VdWHam}
\end{equation}
where the Foerster defect $\Delta_F = E_m - E_i$ is the energy difference between the intermediate and initial state, and $\delta(\omega)$ the Dirac function. Aiming for a simpler notation, the two different Rydberg states can be identified as spin states $\ket{\uparrow}$ and $\ket{\downarrow}$. In the pair state basis $\left( \ket{\uparrow\uparrow}, \ket{\uparrow\downarrow}, \ket{\downarrow\uparrow}, \ket{\downarrow\downarrow}\right)$, the total Hamiltonian describing the interaction between two atoms $i$ and $j$ can be written in matrix form as 
\begin{equation}
\hat{H}_{i,j}^{tot} =
\begin{pmatrix}
E_{\uparrow\uparrow} & 0 & 0 & 0 \\
0 & E_{\downarrow\uparrow} & \frac{J_{\text{ex}}}{2} & 0 \\
0 & \frac{J_{\text{ex}}}{2} & E_{\downarrow\uparrow} & 0 \\
0 & 0 & 0 & E_{\downarrow\downarrow}
\end{pmatrix}.
\end{equation}
with the matrix elements $E_{\uparrow\uparrow} = \bra{\uparrow\uparrow}H_{\text{vdW}}\ket{\uparrow\uparrow}$, $E_{\downarrow\uparrow} = \bra{\downarrow\uparrow}H_{\text{vdW}}\ket{\downarrow\uparrow}$, $E_{\downarrow\downarrow} = \bra{\downarrow\downarrow}H_{\text{vdW}}\ket{\downarrow\downarrow}$ and $J_{\text{ex}} = \bra{\uparrow\downarrow}H_{\text{vdW}}\ket{\downarrow\uparrow}$. This Hamiltonian can be identified as the Heisenberg XXZ Hamiltonian 
\begin{equation}
\begin{aligned}
H_\mathrm{XXZ} = &\frac{1}{2}\sum_{i,j} J_{ij}( S^{(i)}_x S^{(j)}_x &+ S^{(i)}_y S^{(j)}_y + \delta S^{(i)}_z S^{(j)}_z)  \nonumber \\
&+ \sum_i \Delta_{\mathrm{vdW}}S_z^{(i)}
\end{aligned}
\end{equation}
where $J_{ij} = 2J_{\text{ex}}$, $\delta = \left(E_{\downarrow\downarrow} + E_{\uparrow\uparrow} - 2 E_{\downarrow\uparrow}\right)/J_{ij}$ and $\Delta_{\mathrm{vdW}} = (E_{\downarrow\downarrow} - E_{\uparrow\uparrow})/2$. The additional single-spin detuning $\Delta_{\mathrm{vdW}}$ is an order of magnitude smaller than the interaction strength $J_{ij}$ and thus negligible.

The matrix elements $E_{\uparrow\uparrow}$, $E_{\downarrow\uparrow}$, $E_{\downarrow\downarrow}$ and $J_{\text{ex}}$ were calculated using the python module \textit{ARC}~\cite{Sibalic2017}. For the Rydberg states $\ket{48S_{\frac{1}{2}}, +\frac{1}{2}}$ and $\ket{49S_{\frac{1}{2}}, +\frac{1}{2}}$ this yields the interaction strength $J_{ij} = C_6/r_{ij}$ with $C_6/(2\pi) = \SI{59}{\giga \hertz .\micro m \tothe{6}}$ and $\delta = -0.73$.

\section{Determination of the magnetization} \label{sec:Appx_Determination_Magnetization}
\label{sec::appxBDeterminationMagnetization}

The magnetization is extracted from population measurements of the Rydberg states after the readout pulse. To vary the phase $\phi$ of this pulse rapidly enough in order to explore the short time dynamics, the microwave field is generating using frequency up-conversion with a radio-frequency field of frequency 400 MHz, offering time resolution of 10 ns. The populations of the Rydberg states are then extracted using electric field ionization~\cite{gallagher_1994}. At the end of the sequence, a strong electric field of $\SI{100}{\volt\per\centi\meter}$ is switched on and the resulting ions are guided towards a multichannel plate detector. To calibrate the detection efficiency, we combine ionization measurements and depletion imaging~\cite{ferreira2020depletion}. We deduce a detection efficiency $\eta=0.173\pm 0.043$ from four different calibration curves.

At time $t$ we access the magnetization $\langle S_\phi \rangle$ by counting after the readout pulse both the population of the $\ket{\uparrow}$ state $N_\uparrow(\phi)$, and the total spin number $N_{\downarrow+\uparrow}$, according to
\begin{equation}
\langle S_\phi \rangle = \frac{N_{\uparrow}(\phi) - N_{\downarrow}(\phi)}{2N_{\uparrow+\downarrow}} = \frac{N_{\uparrow}(\phi)}{N_{\uparrow+\downarrow}} - \frac{1}{2}.
\label{eq:S_phi_reconstructed}
\end{equation}
Since the ionization is not state-selective, $N_\uparrow(\phi)$ is inferred by counting the spin number after depopulating the $\ket{\downarrow}$ state. It is performed by optically coupling the $\ket{\downarrow}$ state to the short-lived intermediate state $\ket{e}$ during $\SI{1.5}{\micro\second}$. 

\begin{figure}[htb!]
	\hspace*{-0.1cm}
	\includegraphics[width=1\columnwidth]{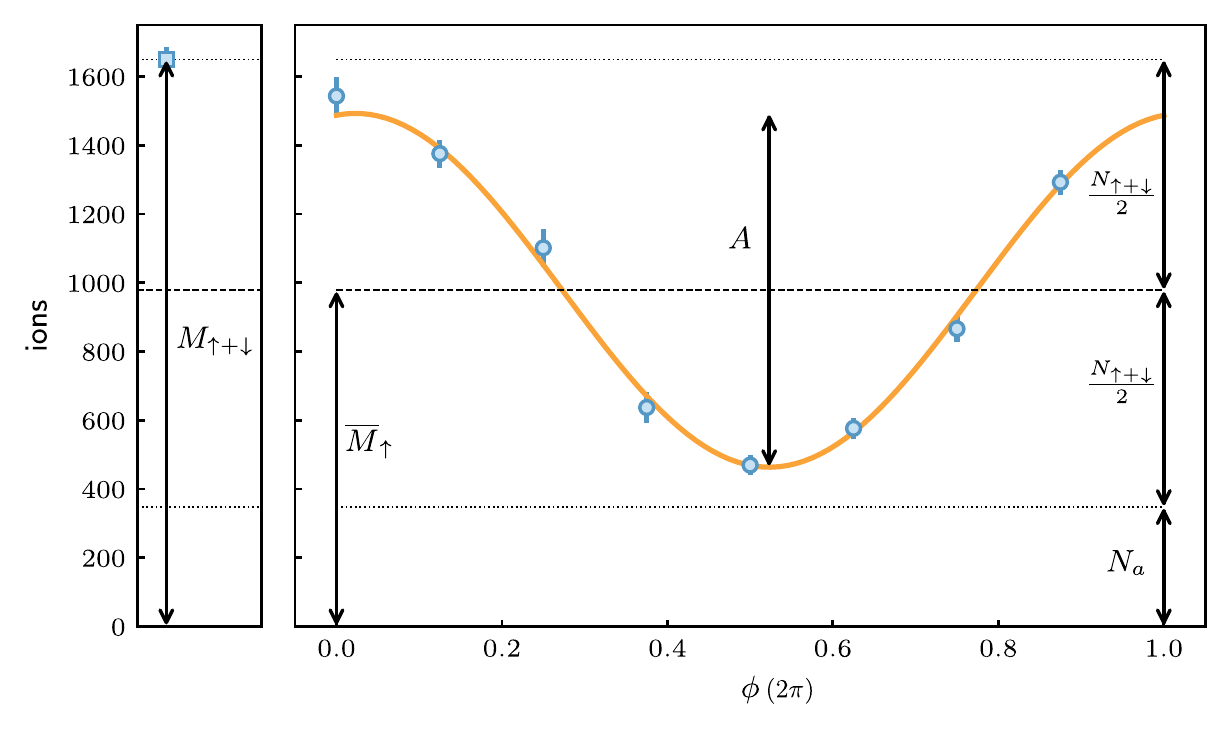}
	\caption{Reconstruction of the magnetization. Left panel: measurements of the total number of spins $M_{\uparrow+\downarrow}$. Right panel: measurement of the population in the $\ket{\uparrow}$ state $M_{\uparrow}(\phi)$ after a readout pulse of phase $\phi$. Both are affected by the population in auxiliary states $N_a$. The measurement  $M_{\uparrow}$ is fitted by a sinusoidal function (orange line), from which we extract the mean value $\overline{M}_{\uparrow}$. The amplitude $A$ of the fit, normalized by the total number of spin $N_{\uparrow+\downarrow}$, indicates the magnetization in the $xy$ plane.
	}
	\label{fig:mag_reconstruction}
\end{figure}

Due to the finite lifetime of the spin states and  microwave transfer, auxiliary Rydberg states might also be populated. This residual population leads to an offset in the measured ion signal, a number $N_a$ of those atoms being energetically above the ionization threshold (see Fig.~\ref{fig:mag_reconstruction}). As a consequence, what we measure instead of $N_{\uparrow}(\phi)$ and $N_{\downarrow+\uparrow}$ are two quantities $M_{\uparrow}(\phi)$ and $M_{\downarrow+\uparrow}$ given by
\begin{align}
M_{\uparrow}(\phi) &= N_{\uparrow}(\phi) + N_a \label{eq:measured_tot}\,, \\
M_{\downarrow+\uparrow} &= N_{\downarrow+\uparrow} + N_a \label{eq:measured_up}\,.
\end{align}
The measured quantity $M_{\uparrow}$ is a sinusoidal function of $\phi$, centered around its mean value
\begin{equation}
\overline{M}_{\uparrow}= \frac{N_{\downarrow+\uparrow}}{2} + N_a\,.
\label{eq:mean_M}
\end{equation}
We determine from a sinusoidal fit the values $M_{\uparrow}(\phi)$ and $\overline{M}_{\uparrow}$ and thus compute the magnetization $\langle S_\phi \rangle$ using~\cref{eq:S_phi_reconstructed,eq:measured_tot,eq:measured_up,eq:mean_M}. The amplitude $A$ of the sinusoidal fit, normalized by $N_{\uparrow+\downarrow}$, corresponds to the magnetization in the $xy$-plane.
Following this procedure, we deduce that the number of atoms in the auxiliary states $N_a$ increases linearly in time with a rate of $\SI{7}{\kilo\hertz}$, consistent with the blackbody decay of the spin states toward Rydberg states above the ionization threshold.

\section{Theoretical models}
\label{sec:appxF_theoretical_models}

To compare the experiment to the mean-field prediction, we solve the classical equations of motion that are obtained from the classical Hamiltonian function~\cite{Schachenmayer2015}
\begin{equation}
H_{\text{C}} = \frac{1}{2}\sum_{i,j} J_{ij}( s^{(i)}_x s^{(j)}_x + s^{(i)}_y s^{(j)}_y + \delta s^{(i)}_z s^{(j)}_z)
\end{equation}
via Hamilton's equation
\begin{equation}
\dot{s}^{(j)}_x = \left\{ s^{(j)}_x, H_{\text{C}} \right\}.
\label{eq:classical_eq_motion}
\end{equation}
Here, $s^{(i)} = (s^{(i)}_{x}, s^{(i)}_{y}, s^{(i)}_{z})$ are classical spins and $\left\{ \dots \right\}$ denotes the Poisson bracket. The system of ordinary differential equations is solved by Tsitouras 5/4 Runge-Kutta method~\cite{Tsitouras2011} using the Julia Differential Equations package~\cite{Rackauckas2017}. For a perfect initial state where all spins are aligned in $x$-direction, mean-field theory does not predict any dynamics. However, the interactions present during the first $\pi/2$-pulse of the Ramsey protocol induce small fluctuations in the initial state. We take these imperfections into account by including the preparation and read-out pulses into the simulations which leads to the dynamics shown by the dotted line in Fig. \ref{fig:time}. For the relevant time-scale of the experiment, these dynamics are negligible.

\begin{figure}[thb!]
	\hspace*{-0.1cm}
	\includegraphics[width=1\columnwidth]{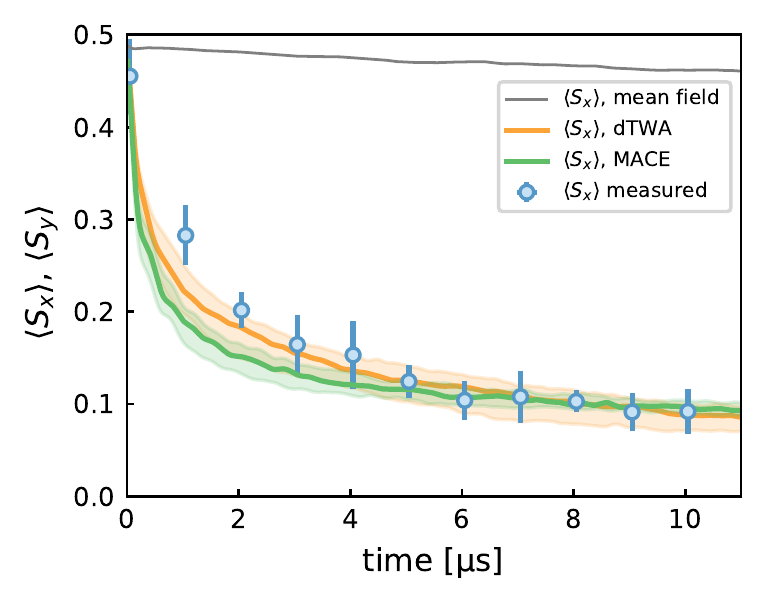}
	\caption{Comparison between dTWA and MACE. Both, MACE and dTWA perform well to describe the experimental data at late times, at intermediate times (between $1$ and \SI{3}{\micro\second}), MACE predicts faster depolarisation dynamics.
	}
	\label{fig:MACE}
\end{figure}

For the considered dynamics, the initial state is an eigenstate of the mean-field Hamiltonian. The relaxation is thus triggered by the initial quantum fluctuations, meaning that mean-field approaches fails in this case. Instead, we use a discrete Truncated Wigner Approximation (dTWA), which still performs classical evolution of the spins following eq.~\eqref{eq:classical_eq_motion} but includes the initial quantum fluctuations into statistical ensembles of the initial state by sampling Monte-Carlo trajectories on discrete phase-space~\cite{Polkovnikov2010,Schachenmayer2015}. For the simulations in this paper, we sample over 100 initial conditions which is sufficient for the magnetization to be converged. Simulating far-from equilibrium dynamics of disordered spin systems has been successfully applied to spin systems in recent work~\cite{Orioli2018}. Imperfections of the preparation and readout are taken into account by simulating the whole Ramsey sequence including those two microwave pulses. We also compared dTWA with an approximate quantum mechanical model, the so-called Moving Average Cluster Expansion~\cite{Hazzard2014} which qualitatively gives similar results (see Fig.~\ref{fig:MACE}).

\section{Quantification of slow dynamics by a stretched exponential} 
\label{sec:appxG_Stretched_Exponential}

\begin{table*}[!htb]
	\setlength\LTleft{0pt}
	\setlength\LTright{0pt}
	\setlength\tabcolsep{0pt}
	\begin{center}
		\begin{footnotesize}
			\begin{longtable}{l|SSSSSSSSS}
				\hline
				\hline
				\multirow{2}{0pt} \textbf{fig.} & {\ $t_{\text{exc}}$} & {$\rho_{\text{gs}}^0$} & {$\rho_{\text{S}}^0$} & {$N_{\text{S}}$} &  {$R_{\text{bl}}$} & {$a$} &  {$\left( \frac{a}{R_{\mathrm{bl}}}\right)^{-3}$} & {$C_6/(2\pi)/a^6$} & {$\beta$} \\*
				&   {[\si{\micro\second}]}  & {$10^{11}$\,[\si{\centi\meter\tothe{-3}}]} & {$10^{9}$\,[\si{\centi\meter\tothe{-3}}]} & {$\times 1000$} & {[\si{\micro\meter}]} & {[\si{\micro\meter}]} & {-} & {[\si{\mega\hertz}]} & {-}   \\
				\hline
				2, 3      &       1.0     &       1.79(9)     &       1.2(3)      &       1.2(3)          &       5.21        &       5.8(5)      &       0.7(2)      &       1.5(8)          &       0.32(2)         \\
				3           &       0.6     &       1.69(12)    &       0.43(11)    &       0.4(1)          &       4.81        &       8.2(7)      &       0.20(5)     &       0.195(97)       &       0.37(2)         \\
				3           &       0.8     &       1.73(9)     &       0.8(2)      &       0.8(2)          &       5.03        &       6.6(6)      &       0.43(11)    &       0.7(3)          &       0.36(4)         \\
				3           &       1.0     &       1.64(15)    &       1.2(3)      &       1.1(3)          &       5.21        &       5.9(5)      &       0.7(2)      &       1.4(7)          &       0.305(14)       \\
				\hline\hline
			\end{longtable}
			\caption{\textbf{Experimental parameters.} $t_{\text{exc}}$ denotes the time of laser excitation from the ground to the Rydberg state. $\rho_{\text{gs}}^0$ denotes the measured ground state density. $\rho_{\text{S}}^0$ the derived peak spin density, $N_{\text{S}}$, the derived number of total spins. $R_{\text{bl}}$ is the blockade radius derived from the excitation time and laser coupling strength. $a$ denotes the Wigner-Seitz radius, $C_6/a^6$ the van der Waals Coefficient	and $\beta$ is the exponent of the stretching exponential derived from a fit to the relaxation curves.}
			\label{Table:Relaxation}
		\end{footnotesize}		
	\end{center}
\end{table*}

A phenomenological approach to describe slow dynamics in disordered systems is a fit of the magnetization with a stretched exponential
\begin{equation}
\langle S_{x}\rangle(t) = \frac{1}{2} \exp(-(\gamma_J t)^{\beta})
\end{equation}
with relaxation rate $\gamma_J$ and stretching exponent $\beta$. This was already proposed by Kohlrausch in 1847~\cite{Kohlrausch1854}, a review on the stretching exponent in numerical simulations and in experimental data of various materials can be found in~\cite{Phillips1996}.\\

For $\beta=1$ the stretched exponential describes an exponential decay. In the limit $\beta \rightarrow 0$, the stretched exponential approaches the logarithmic decay which can be seen by performing a Taylor expansion at small $\beta$
\begin{equation}
\exp(-(t/\tau)^{\beta}) = \frac{1}{e} - \frac{\beta\log(\frac{t}{\tau})}{e} + \mathcal{O}(\beta^3).
\end{equation}
So, the stretching exponent $\beta$ quantifies how slow a system relaxes: A small value signifies that the dynamics are close to logarithmic and slow, a large value indicates fast dynamics.  

The magnetization at early times can be calculated by Baker-Campbell-Haussdorff formula
\begin{equation}
\begin{aligned}
\langle S_x(t) \rangle &= \langle e^{iHt}S_xe^{-iHt}\rangle \\
&= \langle S_x \rangle + it \langle [H, S_x] \rangle - t^2/2\langle[H, [H, S_x]]\rangle + \dots.
\end{aligned}
\end{equation}
Since the initial state $\ket{\rightarrow}_x^{\otimes N}$ is an eigenstate of $S_x$, the expectation value of the commutator $[H, S_x]$ vanishes for this state and we expect the initial dynamics to be quadratic in time. However, this doesn't hold for the stretched exponential function that is a power law with exponent $\beta$ for short times $t \ll \frac{1}{\gamma_J}$:
\begin{align}
\langle S_x\rangle (t) = 1/2\left(1 - \beta\left(\gamma_J t\right)^{\beta}\right)
\end{align}
Therefore, we exclude the very early dynamics from the fit where $t < 1/J_{\max}$ (see Fig.~\ref{fig:hom-rel}).

\section{Spatial spin distribution and disorder strength} 
\label{sec:appxD_spatial_distribution}

To model the experimental 3D spin distribution, we employ a simplified description of the Rydberg excitation dynamics in a cloud of ground-state atoms. Although the experimental procedure creates a superposition of different configurations of atoms being excited to the Rydberg state, each configuration of this superposition can be regarded as an independent disorder realization. Indeed, the different configurations evolve independently from each other under the spin dynamics, and the final projective measurement randomly selects one of them (see appendix~\ref{sec::appxBDeterminationMagnetization}). 
Thus, to create samples of such configurations of Rydberg excitations we iteratively select atoms randomly and excite them to the Rydberg state with a certain excitation probability which we set to zero if another atom within a distance of $R_{\mathrm{bl}}$ is already in the Rydberg state. The excitation probability includes a collective enhancement factor caused by the Rydberg blockade effect~\cite{Weimer2008, Garttner2012}. We also take into account the profile of the laser excitation, characterized by a Gaussian distribution of the two-photon Rabi frequency with measured radius $\sigma=70.6 (3)$ $\mu$m ($e^{-1/2}$), and the Gaussian density distribution of the ground-state atomic cloud (measured radii at $e^{-1/2}$: $\sigma_x=203 (3)$ $\mu$m, $\sigma_y=\sigma_z=35 (1)$ $\mu$m). In our simulations, the peak two-photon Rabi frequency was chosen such that the total number of excited atoms equals the one measured by field ionization.

\begin{figure}[htb!]
	\hspace*{-0.1cm}
	\includegraphics[width=1\columnwidth]{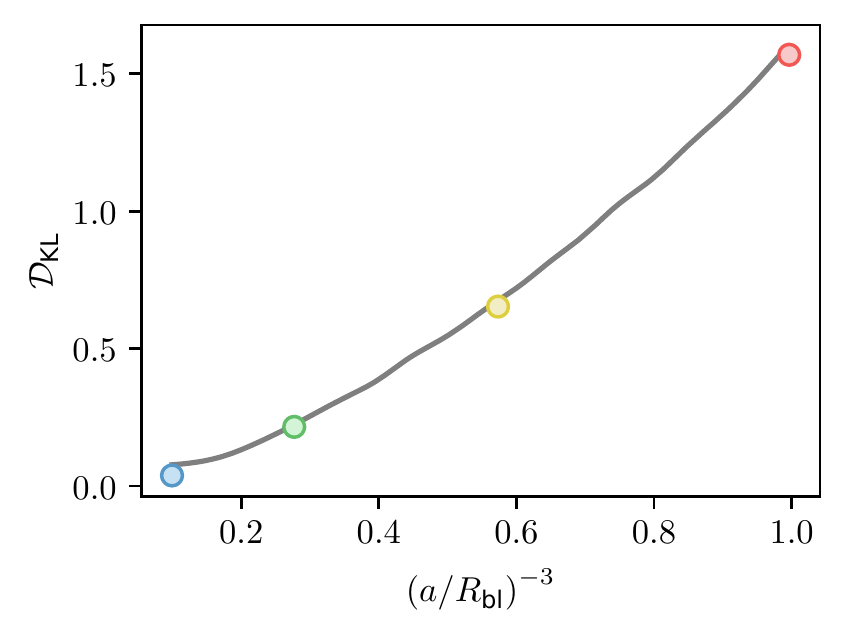}
	\caption{Scaling of the disorder strength as a function of density. The Kullback-Leibler divergence $\mathcal{D}_{\text{KL}}$ increases monotonically with $(a/R_{\text{bl}})^{-3}$. This quantifies the additional correlations induced by the Rydberg blockade effect. The dots indicate the Kullback-Leibler divergence corresponding to the specific simulations presented in Fig.\ref{fig:hom-rel}.
	}
	\label{fig:D_KL_vs_density_large}
\end{figure}

For simulations of a homogeneous system, the spins are randomly distributed in a uniform box taking into account a blockade radius of \SI{5}{\micro\meter} until the desired density$\rho_{\mathrm{S}}$ is reached. In the limit of no blockade effect, the nearest-neighbor distribution for a given Wigner-Seitz radius $a = \left(4\pi \rho/3\right)^{-1/3}$ would be given by~\cite{Hertz1909}
\begin{equation}
h(r) = \frac{3}{a}\left( r/a \right)^2 \exp\left(-\left( r/a \right)^3\right)\,,
\end{equation}
yielding the distribution of coupling strengths $h(J) = h(r)\frac{\partial r}{\partial J}$. Instead, the blockade effect modifies the nearest-neighbour distribution, resulting in a different coupling distribution $g(J)$  (see Fig.~\ref{fig:DKL}~(a)). We quantify the disorder strength of the spin system with the Kullback-Leibler divergence~\cite{Kullback1951}
\begin{equation}
\mathcal{D}_{\text{KL}}(g \parallel h) = \int g(J) \log\left(\frac{g(J)}{h(J)}\right) \mathrm{d}J, 
\end{equation}
i.e. the amount of information that is gained by updating from the distribution $h(J)$. Indeed, the Kulback-Leibler divergence increases almost linearly with density (see Fig.~\ref{fig:D_KL_vs_density_large}). This confirms that $(a/R_{\text{bl}})^{-3}$ is a relevant scale to describe the disorder strength. 

When investigating the universal character of the spin dynamics for homogeneous system, we have concluded that the typical energy scale should be determined by the median of the mean-field energy $J_{\text{mf}} = \text{median}(C_6/\tilde{r}_i^6)$, with $\tilde{r}_i^{-6} = \sum_j r_{ij}^{-6}$ (see Fig.~\ref{fig:hom-rel}). The effective distance $\tilde{a}$, defined by $J_{\text{mf}} = C_6/\tilde{a}^6$, thus corresponds to the median of the non-Gaussian distribution $\{\tilde{r}_i\}$.

\section{Summary of parameters}

Table \ref{Table:Relaxation} summarizes the parameters of the individual measurements shown in figure 2 and 3 of the main text.

\nocite{}
\bibliography{ureldyn_bibliography}

\begin{thebibliography}{62}%
\makeatletter
\providecommand \@ifxundefined [1]{%
 \@ifx{#1\undefined}
}%
\providecommand \@ifnum [1]{%
 \ifnum #1\expandafter \@firstoftwo
 \else \expandafter \@secondoftwo
 \fi
}%
\providecommand \@ifx [1]{%
 \ifx #1\expandafter \@firstoftwo
 \else \expandafter \@secondoftwo
 \fi
}%
\providecommand \natexlab [1]{#1}%
\providecommand \enquote  [1]{``#1''}%
\providecommand \bibnamefont  [1]{#1}%
\providecommand \bibfnamefont [1]{#1}%
\providecommand \citenamefont [1]{#1}%
\providecommand \href@noop [0]{\@secondoftwo}%
\providecommand \href [0]{\begingroup \@sanitize@url \@href}%
\providecommand \@href[1]{\@@startlink{#1}\@@href}%
\providecommand \@@href[1]{\endgroup#1\@@endlink}%
\providecommand \@sanitize@url [0]{\catcode `\\12\catcode `\$12\catcode
  `\&12\catcode `\#12\catcode `\^12\catcode `\_12\catcode `\%12\relax}%
\providecommand \@@startlink[1]{}%
\providecommand \@@endlink[0]{}%
\providecommand \url  [0]{\begingroup\@sanitize@url \@url }%
\providecommand \@url [1]{\endgroup\@href {#1}{\urlprefix }}%
\providecommand \urlprefix  [0]{URL }%
\providecommand \Eprint [0]{\href }%
\providecommand \doibase [0]{https://doi.org/}%
\providecommand \selectlanguage [0]{\@gobble}%
\providecommand \bibinfo  [0]{\@secondoftwo}%
\providecommand \bibfield  [0]{\@secondoftwo}%
\providecommand \translation [1]{[#1]}%
\providecommand \BibitemOpen [0]{}%
\providecommand \bibitemStop [0]{}%
\providecommand \bibitemNoStop [0]{.\EOS\space}%
\providecommand \EOS [0]{\spacefactor3000\relax}%
\providecommand \BibitemShut  [1]{\csname bibitem#1\endcsname}%
\let\auto@bib@innerbib\@empty
\bibitem [{\citenamefont {Rigol}\ \emph {et~al.}(2008)\citenamefont {Rigol},
  \citenamefont {Dunjko},\ and\ \citenamefont {Olshanii}}]{Rigol2008}%
  \BibitemOpen
  \bibfield  {author} {\bibinfo {author} {\bibfnamefont {M.}~\bibnamefont
  {Rigol}}, \bibinfo {author} {\bibfnamefont {V.}~\bibnamefont {Dunjko}},\ and\
  \bibinfo {author} {\bibfnamefont {M.}~\bibnamefont {Olshanii}},\ }\bibfield
  {title} {\bibinfo {title} {{Thermalization and its mechanism for generic
  isolated quantum systems}},\ }\href {https://doi.org/10.1038/nature06838}
  {\bibfield  {journal} {\bibinfo  {journal} {Nature}\ }\textbf {\bibinfo
  {volume} {452}},\ \bibinfo {pages} {854} (\bibinfo {year}
  {2008})}\BibitemShut {NoStop}%
\bibitem [{\citenamefont {Pr{\"{u}}fer}\ \emph {et~al.}(2018)\citenamefont
  {Pr{\"{u}}fer}, \citenamefont {Kunkel}, \citenamefont {Strobel},
  \citenamefont {Lannig}, \citenamefont {Linnemann}, \citenamefont {Schmied},
  \citenamefont {Berges}, \citenamefont {Gasenzer},\ and\ \citenamefont
  {Oberthaler}}]{Prufer2018}%
  \BibitemOpen
  \bibfield  {author} {\bibinfo {author} {\bibfnamefont {M.}~\bibnamefont
  {Pr{\"{u}}fer}}, \bibinfo {author} {\bibfnamefont {P.}~\bibnamefont
  {Kunkel}}, \bibinfo {author} {\bibfnamefont {H.}~\bibnamefont {Strobel}},
  \bibinfo {author} {\bibfnamefont {S.}~\bibnamefont {Lannig}}, \bibinfo
  {author} {\bibfnamefont {D.}~\bibnamefont {Linnemann}}, \bibinfo {author}
  {\bibfnamefont {C.~M.}\ \bibnamefont {Schmied}}, \bibinfo {author}
  {\bibfnamefont {J.}~\bibnamefont {Berges}}, \bibinfo {author} {\bibfnamefont
  {T.}~\bibnamefont {Gasenzer}},\ and\ \bibinfo {author} {\bibfnamefont
  {M.~K.}\ \bibnamefont {Oberthaler}},\ }\bibfield  {title} {\bibinfo {title}
  {{Observation of universal dynamics in a spinor Bose gas far from
  equilibrium}},\ }\href {https://doi.org/10.1038/s41586-018-0659-0} {\bibfield
   {journal} {\bibinfo  {journal} {Nature}\ }\textbf {\bibinfo {volume}
  {563}},\ \bibinfo {pages} {217} (\bibinfo {year} {2018})}\BibitemShut
  {NoStop}%
\bibitem [{\citenamefont {Eigen}\ \emph {et~al.}(2018)\citenamefont {Eigen},
  \citenamefont {Glidden}, \citenamefont {Lopes}, \citenamefont {Cornell},
  \citenamefont {Smith},\ and\ \citenamefont {Hadzibabic}}]{Eigen2018}%
  \BibitemOpen
  \bibfield  {author} {\bibinfo {author} {\bibfnamefont {C.}~\bibnamefont
  {Eigen}}, \bibinfo {author} {\bibfnamefont {J.~A.}\ \bibnamefont {Glidden}},
  \bibinfo {author} {\bibfnamefont {R.}~\bibnamefont {Lopes}}, \bibinfo
  {author} {\bibfnamefont {E.~A.}\ \bibnamefont {Cornell}}, \bibinfo {author}
  {\bibfnamefont {R.~P.}\ \bibnamefont {Smith}},\ and\ \bibinfo {author}
  {\bibfnamefont {Z.}~\bibnamefont {Hadzibabic}},\ }\bibfield  {title}
  {\bibinfo {title} {{Universal prethermal dynamics of Bose gases quenched to
  unitarity}},\ }\href {https://doi.org/10.1038/s41586-018-0674-1} {\bibfield
  {journal} {\bibinfo  {journal} {Nature}\ }\textbf {\bibinfo {volume} {563}},\
  \bibinfo {pages} {221} (\bibinfo {year} {2018})}\BibitemShut {NoStop}%
\bibitem [{\citenamefont {Erne}\ \emph {et~al.}(2018)\citenamefont {Erne},
  \citenamefont {B{\"{u}}cker}, \citenamefont {Gasenzer}, \citenamefont
  {Berges},\ and\ \citenamefont {Schmiedmayer}}]{Erne2018}%
  \BibitemOpen
  \bibfield  {author} {\bibinfo {author} {\bibfnamefont {S.}~\bibnamefont
  {Erne}}, \bibinfo {author} {\bibfnamefont {R.}~\bibnamefont {B{\"{u}}cker}},
  \bibinfo {author} {\bibfnamefont {T.}~\bibnamefont {Gasenzer}}, \bibinfo
  {author} {\bibfnamefont {J.}~\bibnamefont {Berges}},\ and\ \bibinfo {author}
  {\bibfnamefont {J.}~\bibnamefont {Schmiedmayer}},\ }\bibfield  {title}
  {\bibinfo {title} {{Universal dynamics in an isolated one-dimensional Bose
  gas far from equilibrium}},\ }\href
  {https://doi.org/10.1038/s41586-018-0667-0} {\bibfield  {journal} {\bibinfo
  {journal} {Nature}\ }\textbf {\bibinfo {volume} {563}},\ \bibinfo {pages}
  {225} (\bibinfo {year} {2018})}\BibitemShut {NoStop}%
\bibitem [{\citenamefont {Yao}\ \emph {et~al.}(2014)\citenamefont {Yao},
  \citenamefont {Laumann}, \citenamefont {Gopalakrishnan}, \citenamefont
  {Knap}, \citenamefont {M{\"{u}}ller}, \citenamefont {Demler},\ and\
  \citenamefont {Lukin}}]{Yao2014}%
  \BibitemOpen
  \bibfield  {author} {\bibinfo {author} {\bibfnamefont {N.~Y.}\ \bibnamefont
  {Yao}}, \bibinfo {author} {\bibfnamefont {C.~R.}\ \bibnamefont {Laumann}},
  \bibinfo {author} {\bibfnamefont {S.}~\bibnamefont {Gopalakrishnan}},
  \bibinfo {author} {\bibfnamefont {M.}~\bibnamefont {Knap}}, \bibinfo {author}
  {\bibfnamefont {M.}~\bibnamefont {M{\"{u}}ller}}, \bibinfo {author}
  {\bibfnamefont {E.~A.}\ \bibnamefont {Demler}},\ and\ \bibinfo {author}
  {\bibfnamefont {M.~D.}\ \bibnamefont {Lukin}},\ }\bibfield  {title} {\bibinfo
  {title} {{Many-body localization in dipolar systems}},\ }\href
  {https://doi.org/10.1103/PhysRevLett.113.243002} {\bibfield  {journal}
  {\bibinfo  {journal} {Physical Review Letters}\ }\textbf {\bibinfo {volume}
  {113}},\ \bibinfo {pages} {243002} (\bibinfo {year} {2014})}\BibitemShut
  {NoStop}%
\bibitem [{\citenamefont {Smith}\ \emph {et~al.}(2016)\citenamefont {Smith},
  \citenamefont {Lee}, \citenamefont {Richerme}, \citenamefont {Neyenhuis},
  \citenamefont {Hess}, \citenamefont {Hauke}, \citenamefont {Heyl},
  \citenamefont {Huse},\ and\ \citenamefont {Monroe}}]{Smith2016}%
  \BibitemOpen
  \bibfield  {author} {\bibinfo {author} {\bibfnamefont {J.}~\bibnamefont
  {Smith}}, \bibinfo {author} {\bibfnamefont {A.}~\bibnamefont {Lee}}, \bibinfo
  {author} {\bibfnamefont {P.}~\bibnamefont {Richerme}}, \bibinfo {author}
  {\bibfnamefont {B.}~\bibnamefont {Neyenhuis}}, \bibinfo {author}
  {\bibfnamefont {P.~W.}\ \bibnamefont {Hess}}, \bibinfo {author}
  {\bibfnamefont {P.}~\bibnamefont {Hauke}}, \bibinfo {author} {\bibfnamefont
  {M.}~\bibnamefont {Heyl}}, \bibinfo {author} {\bibfnamefont {D.~A.}\
  \bibnamefont {Huse}},\ and\ \bibinfo {author} {\bibfnamefont
  {C.}~\bibnamefont {Monroe}},\ }\bibfield  {title} {\bibinfo {title}
  {{Many-body localization in a quantum simulator with programmable random
  disorder}},\ }\href {https://doi.org/10.1038/nphys3783} {\bibfield  {journal}
  {\bibinfo  {journal} {Nature Physics}\ }\textbf {\bibinfo {volume} {12}},\
  \bibinfo {pages} {907} (\bibinfo {year} {2016})}\BibitemShut {NoStop}%
\bibitem [{\citenamefont {Choi}\ \emph {et~al.}(2016)\citenamefont {Choi},
  \citenamefont {Hild}, \citenamefont {Zeiher}, \citenamefont {Schau{\ss}},
  \citenamefont {Rubio-Abadal}, \citenamefont {Yefsah}, \citenamefont
  {Khemani}, \citenamefont {Huse}, \citenamefont {Bloch},\ and\ \citenamefont
  {Gross}}]{Choi2016}%
  \BibitemOpen
  \bibfield  {author} {\bibinfo {author} {\bibfnamefont {J.~Y.}\ \bibnamefont
  {Choi}}, \bibinfo {author} {\bibfnamefont {S.}~\bibnamefont {Hild}}, \bibinfo
  {author} {\bibfnamefont {J.}~\bibnamefont {Zeiher}}, \bibinfo {author}
  {\bibfnamefont {P.}~\bibnamefont {Schau{\ss}}}, \bibinfo {author}
  {\bibfnamefont {A.}~\bibnamefont {Rubio-Abadal}}, \bibinfo {author}
  {\bibfnamefont {T.}~\bibnamefont {Yefsah}}, \bibinfo {author} {\bibfnamefont
  {V.}~\bibnamefont {Khemani}}, \bibinfo {author} {\bibfnamefont {D.~A.}\
  \bibnamefont {Huse}}, \bibinfo {author} {\bibfnamefont {I.}~\bibnamefont
  {Bloch}},\ and\ \bibinfo {author} {\bibfnamefont {C.}~\bibnamefont {Gross}},\
  }\bibfield  {title} {\bibinfo {title} {{Exploring the many-body localization
  transition in two dimensions}},\ }\href
  {https://doi.org/10.1126/science.aaf8834} {\bibfield  {journal} {\bibinfo
  {journal} {Science}\ }\textbf {\bibinfo {volume} {352}},\ \bibinfo {pages}
  {1547} (\bibinfo {year} {2016})}\BibitemShut {NoStop}%
\bibitem [{\citenamefont {Choi}\ \emph
  {et~al.}(2017{\natexlab{a}})\citenamefont {Choi}, \citenamefont {Choi},
  \citenamefont {Landig}, \citenamefont {Kucsko}, \citenamefont {Zhou},
  \citenamefont {Isoya}, \citenamefont {Jelezko}, \citenamefont {Onoda},
  \citenamefont {Sumiya}, \citenamefont {Khemani}, \citenamefont {von
  Keyserlingk}, \citenamefont {Yao}, \citenamefont {Demler},\ and\
  \citenamefont {Lukin}}]{Choi2017_tc}%
  \BibitemOpen
  \bibfield  {author} {\bibinfo {author} {\bibfnamefont {S.}~\bibnamefont
  {Choi}}, \bibinfo {author} {\bibfnamefont {J.}~\bibnamefont {Choi}}, \bibinfo
  {author} {\bibfnamefont {R.}~\bibnamefont {Landig}}, \bibinfo {author}
  {\bibfnamefont {G.}~\bibnamefont {Kucsko}}, \bibinfo {author} {\bibfnamefont
  {H.}~\bibnamefont {Zhou}}, \bibinfo {author} {\bibfnamefont {J.}~\bibnamefont
  {Isoya}}, \bibinfo {author} {\bibfnamefont {F.}~\bibnamefont {Jelezko}},
  \bibinfo {author} {\bibfnamefont {S.}~\bibnamefont {Onoda}}, \bibinfo
  {author} {\bibfnamefont {H.}~\bibnamefont {Sumiya}}, \bibinfo {author}
  {\bibfnamefont {V.}~\bibnamefont {Khemani}}, \bibinfo {author} {\bibfnamefont
  {C.}~\bibnamefont {von Keyserlingk}}, \bibinfo {author} {\bibfnamefont
  {N.~Y.}\ \bibnamefont {Yao}}, \bibinfo {author} {\bibfnamefont
  {E.}~\bibnamefont {Demler}},\ and\ \bibinfo {author} {\bibfnamefont {M.~D.}\
  \bibnamefont {Lukin}},\ }\bibfield  {title} {\bibinfo {title} {{Observation
  of discrete time-crystalline order in a disordered dipolar many-body
  system}},\ }\href {http://dx.doi.org/10.1038/nature21426
  http://10.0.4.14/nature21426} {\bibfield  {journal} {\bibinfo  {journal}
  {Nature}\ }\textbf {\bibinfo {volume} {543}},\ \bibinfo {pages} {221}
  (\bibinfo {year} {2017}{\natexlab{a}})}\BibitemShut {NoStop}%
\bibitem [{\citenamefont {Zhang}\ \emph {et~al.}(2017)\citenamefont {Zhang},
  \citenamefont {Hess}, \citenamefont {Kyprianidis}, \citenamefont {Becker},
  \citenamefont {Lee}, \citenamefont {Smith}, \citenamefont {Pagano},
  \citenamefont {Potirniche}, \citenamefont {Potter}, \citenamefont
  {Vishwanath}, \citenamefont {Yao},\ and\ \citenamefont {Monroe}}]{Zhang2017}%
  \BibitemOpen
  \bibfield  {author} {\bibinfo {author} {\bibfnamefont {J.}~\bibnamefont
  {Zhang}}, \bibinfo {author} {\bibfnamefont {P.~W.}\ \bibnamefont {Hess}},
  \bibinfo {author} {\bibfnamefont {A.}~\bibnamefont {Kyprianidis}}, \bibinfo
  {author} {\bibfnamefont {P.}~\bibnamefont {Becker}}, \bibinfo {author}
  {\bibfnamefont {A.}~\bibnamefont {Lee}}, \bibinfo {author} {\bibfnamefont
  {J.}~\bibnamefont {Smith}}, \bibinfo {author} {\bibfnamefont
  {G.}~\bibnamefont {Pagano}}, \bibinfo {author} {\bibfnamefont {I.-D.}\
  \bibnamefont {Potirniche}}, \bibinfo {author} {\bibfnamefont {A.~C.}\
  \bibnamefont {Potter}}, \bibinfo {author} {\bibfnamefont {A.}~\bibnamefont
  {Vishwanath}}, \bibinfo {author} {\bibfnamefont {N.~Y.}\ \bibnamefont
  {Yao}},\ and\ \bibinfo {author} {\bibfnamefont {C.}~\bibnamefont {Monroe}},\
  }\bibfield  {title} {\bibinfo {title} {{Observation of a discrete time
  crystal}},\ }\href {http://dx.doi.org/10.1038/nature21413
  http://10.0.4.14/nature21413} {\bibfield  {journal} {\bibinfo  {journal}
  {Nature}\ }\textbf {\bibinfo {volume} {543}},\ \bibinfo {pages} {217}
  (\bibinfo {year} {2017})}\BibitemShut {NoStop}%
\bibitem [{\citenamefont {Bernien}\ \emph {et~al.}(2017)\citenamefont
  {Bernien}, \citenamefont {Schwartz}, \citenamefont {Keesling}, \citenamefont
  {Levine}, \citenamefont {Omran}, \citenamefont {Pichler}, \citenamefont
  {Choi}, \citenamefont {Zibrov}, \citenamefont {Endres}, \citenamefont
  {Greiner}, \citenamefont {Vuletic},\ and\ \citenamefont
  {Lukin}}]{Bernien2017}%
  \BibitemOpen
  \bibfield  {author} {\bibinfo {author} {\bibfnamefont {H.}~\bibnamefont
  {Bernien}}, \bibinfo {author} {\bibfnamefont {S.}~\bibnamefont {Schwartz}},
  \bibinfo {author} {\bibfnamefont {A.}~\bibnamefont {Keesling}}, \bibinfo
  {author} {\bibfnamefont {H.}~\bibnamefont {Levine}}, \bibinfo {author}
  {\bibfnamefont {A.}~\bibnamefont {Omran}}, \bibinfo {author} {\bibfnamefont
  {H.}~\bibnamefont {Pichler}}, \bibinfo {author} {\bibfnamefont
  {S.}~\bibnamefont {Choi}}, \bibinfo {author} {\bibfnamefont {A.~S.}\
  \bibnamefont {Zibrov}}, \bibinfo {author} {\bibfnamefont {M.}~\bibnamefont
  {Endres}}, \bibinfo {author} {\bibfnamefont {M.}~\bibnamefont {Greiner}},
  \bibinfo {author} {\bibfnamefont {V.}~\bibnamefont {Vuletic}},\ and\ \bibinfo
  {author} {\bibfnamefont {M.~D.}\ \bibnamefont {Lukin}},\ }\bibfield  {title}
  {\bibinfo {title} {{Probing many-body dynamics on a 51-atom quantum
  simulator}},\ }\href {https://doi.org/10.1038/nature24622} {\bibfield
  {journal} {\bibinfo  {journal} {Nature}\ }\textbf {\bibinfo {volume} {551}},\
  \bibinfo {pages} {579} (\bibinfo {year} {2017})},\ \Eprint
  {https://arxiv.org/abs/1707.04344} {arXiv:1707.04344} \BibitemShut {NoStop}%
\bibitem [{\citenamefont {Turner}\ \emph {et~al.}(2018)\citenamefont {Turner},
  \citenamefont {Michailidis}, \citenamefont {Abanin}, \citenamefont {Serbyn},\
  and\ \citenamefont {Papi{\'{c}}}}]{Turner2018}%
  \BibitemOpen
  \bibfield  {author} {\bibinfo {author} {\bibfnamefont {C.~J.}\ \bibnamefont
  {Turner}}, \bibinfo {author} {\bibfnamefont {A.~A.}\ \bibnamefont
  {Michailidis}}, \bibinfo {author} {\bibfnamefont {D.~A.}\ \bibnamefont
  {Abanin}}, \bibinfo {author} {\bibfnamefont {M.}~\bibnamefont {Serbyn}},\
  and\ \bibinfo {author} {\bibfnamefont {Z.}~\bibnamefont {Papi{\'{c}}}},\
  }\bibfield  {title} {\bibinfo {title} {{Weak ergodicity breaking from quantum
  many-body scars}},\ }\href {https://doi.org/10.1038/s41567-018-0137-5}
  {\bibfield  {journal} {\bibinfo  {journal} {Nature Physics}\ }\textbf
  {\bibinfo {volume} {14}},\ \bibinfo {pages} {745} (\bibinfo {year}
  {2018})}\BibitemShut {NoStop}%
\bibitem [{\citenamefont {Parameswaran}\ \emph {et~al.}(2017)\citenamefont
  {Parameswaran}, \citenamefont {Potter},\ and\ \citenamefont
  {Vasseur}}]{Parameswaran2017}%
  \BibitemOpen
  \bibfield  {author} {\bibinfo {author} {\bibfnamefont {S.~A.}\ \bibnamefont
  {Parameswaran}}, \bibinfo {author} {\bibfnamefont {A.~C.}\ \bibnamefont
  {Potter}},\ and\ \bibinfo {author} {\bibfnamefont {R.}~\bibnamefont
  {Vasseur}},\ }\bibfield  {title} {\bibinfo {title} {{Eigenstate phase
  transitions and the emergence of universal dynamics in highly excited
  states}},\ }\href {https://doi.org/10.1002/andp.201600302} {\bibfield
  {journal} {\bibinfo  {journal} {Annalen der Physik}\ }\textbf {\bibinfo
  {volume} {529}},\ \bibinfo {pages} {1600302} (\bibinfo {year}
  {2017})}\BibitemShut {NoStop}%
\bibitem [{\citenamefont {Binder}\ and\ \citenamefont
  {Young}(1986)}]{Binder1986}%
  \BibitemOpen
  \bibfield  {author} {\bibinfo {author} {\bibfnamefont {K.}~\bibnamefont
  {Binder}}\ and\ \bibinfo {author} {\bibfnamefont {A.~P.}\ \bibnamefont
  {Young}},\ }\bibfield  {title} {\bibinfo {title} {{Spin glasses: Experimental
  facts, theoretical concepts, and open questions}},\ }\href
  {https://doi.org/10.1103/RevModPhys.58.801} {\bibfield  {journal} {\bibinfo
  {journal} {Reviews of Modern Physics}\ }\textbf {\bibinfo {volume} {58}},\
  \bibinfo {pages} {801} (\bibinfo {year} {1986})}\BibitemShut {NoStop}%
\bibitem [{\citenamefont {Ladizinsky}\ \emph {et~al.}(2018)\citenamefont
  {Ladizinsky}, \citenamefont {Ladizinsky}, \citenamefont {Medina},
  \citenamefont {Altomare}, \citenamefont {Enderud}, \citenamefont {Johnson},
  \citenamefont {Yao}, \citenamefont {Hoskinson}, \citenamefont {Sato},
  \citenamefont {Smirnov}, \citenamefont {Li}, \citenamefont {Tsai},
  \citenamefont {Berkley}, \citenamefont {Swenson}, \citenamefont {Bunyk},
  \citenamefont {Huang}, \citenamefont {Rich}, \citenamefont {Deng},
  \citenamefont {Molavi}, \citenamefont {Boothby}, \citenamefont {Whittaker},
  \citenamefont {Neufeld}, \citenamefont {Poulin-Lamarre}, \citenamefont {Oh},
  \citenamefont {Amin}, \citenamefont {Pavlov}, \citenamefont {Perminov},
  \citenamefont {Reis}, \citenamefont {Lanting}, \citenamefont {Harris},\ and\
  \citenamefont {Volkmann}}]{Harris2018}%
  \BibitemOpen
  \bibfield  {author} {\bibinfo {author} {\bibfnamefont {E.}~\bibnamefont
  {Ladizinsky}}, \bibinfo {author} {\bibfnamefont {N.}~\bibnamefont
  {Ladizinsky}}, \bibinfo {author} {\bibfnamefont {T.}~\bibnamefont {Medina}},
  \bibinfo {author} {\bibfnamefont {F.}~\bibnamefont {Altomare}}, \bibinfo
  {author} {\bibfnamefont {C.}~\bibnamefont {Enderud}}, \bibinfo {author}
  {\bibfnamefont {M.~W.}\ \bibnamefont {Johnson}}, \bibinfo {author}
  {\bibfnamefont {J.}~\bibnamefont {Yao}}, \bibinfo {author} {\bibfnamefont
  {E.}~\bibnamefont {Hoskinson}}, \bibinfo {author} {\bibfnamefont
  {Y.}~\bibnamefont {Sato}}, \bibinfo {author} {\bibfnamefont {A.}~\bibnamefont
  {Smirnov}}, \bibinfo {author} {\bibfnamefont {R.}~\bibnamefont {Li}},
  \bibinfo {author} {\bibfnamefont {N.}~\bibnamefont {Tsai}}, \bibinfo {author}
  {\bibfnamefont {A.~J.}\ \bibnamefont {Berkley}}, \bibinfo {author}
  {\bibfnamefont {L.}~\bibnamefont {Swenson}}, \bibinfo {author} {\bibfnamefont
  {P.}~\bibnamefont {Bunyk}}, \bibinfo {author} {\bibfnamefont
  {S.}~\bibnamefont {Huang}}, \bibinfo {author} {\bibfnamefont
  {C.}~\bibnamefont {Rich}}, \bibinfo {author} {\bibfnamefont {C.}~\bibnamefont
  {Deng}}, \bibinfo {author} {\bibfnamefont {R.}~\bibnamefont {Molavi}},
  \bibinfo {author} {\bibfnamefont {K.}~\bibnamefont {Boothby}}, \bibinfo
  {author} {\bibfnamefont {J.}~\bibnamefont {Whittaker}}, \bibinfo {author}
  {\bibfnamefont {R.}~\bibnamefont {Neufeld}}, \bibinfo {author} {\bibfnamefont
  {G.}~\bibnamefont {Poulin-Lamarre}}, \bibinfo {author} {\bibfnamefont
  {T.}~\bibnamefont {Oh}}, \bibinfo {author} {\bibfnamefont {M.~H.}\
  \bibnamefont {Amin}}, \bibinfo {author} {\bibfnamefont {I.}~\bibnamefont
  {Pavlov}}, \bibinfo {author} {\bibfnamefont {I.}~\bibnamefont {Perminov}},
  \bibinfo {author} {\bibfnamefont {M.}~\bibnamefont {Reis}}, \bibinfo {author}
  {\bibfnamefont {T.}~\bibnamefont {Lanting}}, \bibinfo {author} {\bibfnamefont
  {R.}~\bibnamefont {Harris}},\ and\ \bibinfo {author} {\bibfnamefont
  {M.}~\bibnamefont {Volkmann}},\ }\bibfield  {title} {\bibinfo {title} {{Phase
  transitions in a programmable quantum spin glass simulator}},\ }\href
  {https://doi.org/10.1126/science.aat2025} {\bibfield  {journal} {\bibinfo
  {journal} {Science}\ }\textbf {\bibinfo {volume} {361}},\ \bibinfo {pages}
  {162} (\bibinfo {year} {2018})}\BibitemShut {NoStop}%
\bibitem [{\citenamefont {Gezo}\ \emph {et~al.}(2013)\citenamefont {Gezo},
  \citenamefont {Lui}, \citenamefont {Wolin}, \citenamefont {Slichter},
  \citenamefont {Giannetta},\ and\ \citenamefont {Schlueter}}]{Gezo2013}%
  \BibitemOpen
  \bibfield  {author} {\bibinfo {author} {\bibfnamefont {J.}~\bibnamefont
  {Gezo}}, \bibinfo {author} {\bibfnamefont {T.~K.}\ \bibnamefont {Lui}},
  \bibinfo {author} {\bibfnamefont {B.}~\bibnamefont {Wolin}}, \bibinfo
  {author} {\bibfnamefont {C.~P.}\ \bibnamefont {Slichter}}, \bibinfo {author}
  {\bibfnamefont {R.}~\bibnamefont {Giannetta}},\ and\ \bibinfo {author}
  {\bibfnamefont {J.~A.}\ \bibnamefont {Schlueter}},\ }\bibfield  {title}
  {\bibinfo {title} {{Stretched exponential spin relaxation in organic
  superconductors}},\ }\href {https://doi.org/10.1103/PhysRevB.88.140504}
  {\bibfield  {journal} {\bibinfo  {journal} {Physical Review B - Condensed
  Matter and Materials Physics}\ }\textbf {\bibinfo {volume} {88}},\ \bibinfo
  {pages} {140504} (\bibinfo {year} {2013})}\BibitemShut {NoStop}%
\bibitem [{\citenamefont {Dzugutov}\ and\ \citenamefont
  {Phillips}(1995)}]{Dzugutov1995}%
  \BibitemOpen
  \bibfield  {author} {\bibinfo {author} {\bibfnamefont {M.}~\bibnamefont
  {Dzugutov}}\ and\ \bibinfo {author} {\bibfnamefont {J.~C.}\ \bibnamefont
  {Phillips}},\ }\bibfield  {title} {\bibinfo {title} {{Structural relaxation
  in an equilibrium quasicrystal}},\ }\href
  {https://doi.org/10.1016/0022-3093(95)00425-4} {\bibfield  {journal}
  {\bibinfo  {journal} {Journal of Non-Crystalline Solids}\ }\textbf {\bibinfo
  {volume} {192-193}},\ \bibinfo {pages} {397} (\bibinfo {year}
  {1995})}\BibitemShut {NoStop}%
\bibitem [{\citenamefont {L{\"{u}}schen}\ \emph {et~al.}(2017)\citenamefont
  {L{\"{u}}schen}, \citenamefont {Bordia}, \citenamefont {Hodgman},
  \citenamefont {Schreiber}, \citenamefont {Sarkar}, \citenamefont {Daley},
  \citenamefont {Fischer}, \citenamefont {Altman}, \citenamefont {Bloch},\ and\
  \citenamefont {Schneider}}]{Luschen2017}%
  \BibitemOpen
  \bibfield  {author} {\bibinfo {author} {\bibfnamefont {H.~P.}\ \bibnamefont
  {L{\"{u}}schen}}, \bibinfo {author} {\bibfnamefont {P.}~\bibnamefont
  {Bordia}}, \bibinfo {author} {\bibfnamefont {S.~S.}\ \bibnamefont {Hodgman}},
  \bibinfo {author} {\bibfnamefont {M.}~\bibnamefont {Schreiber}}, \bibinfo
  {author} {\bibfnamefont {S.}~\bibnamefont {Sarkar}}, \bibinfo {author}
  {\bibfnamefont {A.~J.}\ \bibnamefont {Daley}}, \bibinfo {author}
  {\bibfnamefont {M.~H.}\ \bibnamefont {Fischer}}, \bibinfo {author}
  {\bibfnamefont {E.}~\bibnamefont {Altman}}, \bibinfo {author} {\bibfnamefont
  {I.}~\bibnamefont {Bloch}},\ and\ \bibinfo {author} {\bibfnamefont
  {U.}~\bibnamefont {Schneider}},\ }\bibfield  {title} {\bibinfo {title}
  {{Signatures of many-body localization in a controlled open quantum
  system}},\ }\href {https://doi.org/10.1103/PhysRevX.7.011034} {\bibfield
  {journal} {\bibinfo  {journal} {Physical Review X}\ }\textbf {\bibinfo
  {volume} {7}},\ \bibinfo {pages} {11034} (\bibinfo {year}
  {2017})}\BibitemShut {NoStop}%
\bibitem [{\citenamefont {Choi}\ \emph
  {et~al.}(2017{\natexlab{b}})\citenamefont {Choi}, \citenamefont {Choi},
  \citenamefont {Kucsko}, \citenamefont {Maurer}, \citenamefont {Shields},
  \citenamefont {Sumiya}, \citenamefont {Onoda}, \citenamefont {Isoya},
  \citenamefont {Demler}, \citenamefont {Jelezko}, \citenamefont {Yao},\ and\
  \citenamefont {Lukin}}]{Choi2017}%
  \BibitemOpen
  \bibfield  {author} {\bibinfo {author} {\bibfnamefont {J.}~\bibnamefont
  {Choi}}, \bibinfo {author} {\bibfnamefont {S.}~\bibnamefont {Choi}}, \bibinfo
  {author} {\bibfnamefont {G.}~\bibnamefont {Kucsko}}, \bibinfo {author}
  {\bibfnamefont {P.~C.}\ \bibnamefont {Maurer}}, \bibinfo {author}
  {\bibfnamefont {B.~J.}\ \bibnamefont {Shields}}, \bibinfo {author}
  {\bibfnamefont {H.}~\bibnamefont {Sumiya}}, \bibinfo {author} {\bibfnamefont
  {S.}~\bibnamefont {Onoda}}, \bibinfo {author} {\bibfnamefont
  {J.}~\bibnamefont {Isoya}}, \bibinfo {author} {\bibfnamefont
  {E.}~\bibnamefont {Demler}}, \bibinfo {author} {\bibfnamefont
  {F.}~\bibnamefont {Jelezko}}, \bibinfo {author} {\bibfnamefont {N.~Y.}\
  \bibnamefont {Yao}},\ and\ \bibinfo {author} {\bibfnamefont {M.~D.}\
  \bibnamefont {Lukin}},\ }\bibfield  {title} {\bibinfo {title}
  {{Depolarization Dynamics in a Strongly Interacting Solid-State Spin
  Ensemble}},\ }\href {https://doi.org/10.1103/PhysRevLett.118.093601}
  {\bibfield  {journal} {\bibinfo  {journal} {Physical Review Letters}\
  }\textbf {\bibinfo {volume} {118}},\ \bibinfo {pages} {093601} (\bibinfo
  {year} {2017}{\natexlab{b}})}\BibitemShut {NoStop}%
\bibitem [{\citenamefont {Kucsko}\ \emph {et~al.}(2018)\citenamefont {Kucsko},
  \citenamefont {Choi}, \citenamefont {Choi}, \citenamefont {Maurer},
  \citenamefont {Zhou}, \citenamefont {Landig}, \citenamefont {Sumiya},
  \citenamefont {Onoda}, \citenamefont {Isoya}, \citenamefont {Jelezko},
  \citenamefont {Demler}, \citenamefont {Yao},\ and\ \citenamefont
  {Lukin}}]{Kucsko2018}%
  \BibitemOpen
  \bibfield  {author} {\bibinfo {author} {\bibfnamefont {G.}~\bibnamefont
  {Kucsko}}, \bibinfo {author} {\bibfnamefont {S.}~\bibnamefont {Choi}},
  \bibinfo {author} {\bibfnamefont {J.}~\bibnamefont {Choi}}, \bibinfo {author}
  {\bibfnamefont {P.~C.}\ \bibnamefont {Maurer}}, \bibinfo {author}
  {\bibfnamefont {H.}~\bibnamefont {Zhou}}, \bibinfo {author} {\bibfnamefont
  {R.}~\bibnamefont {Landig}}, \bibinfo {author} {\bibfnamefont
  {H.}~\bibnamefont {Sumiya}}, \bibinfo {author} {\bibfnamefont
  {S.}~\bibnamefont {Onoda}}, \bibinfo {author} {\bibfnamefont
  {J.}~\bibnamefont {Isoya}}, \bibinfo {author} {\bibfnamefont
  {F.}~\bibnamefont {Jelezko}}, \bibinfo {author} {\bibfnamefont
  {E.}~\bibnamefont {Demler}}, \bibinfo {author} {\bibfnamefont {N.~Y.}\
  \bibnamefont {Yao}},\ and\ \bibinfo {author} {\bibfnamefont {M.~D.}\
  \bibnamefont {Lukin}},\ }\bibfield  {title} {\bibinfo {title} {{Critical
  Thermalization of a Disordered Dipolar Spin System in Diamond}},\ }\href
  {https://doi.org/10.1103/PhysRevLett.121.023601} {\bibfield  {journal}
  {\bibinfo  {journal} {Physical Review Letters}\ }\textbf {\bibinfo {volume}
  {121}},\ \bibinfo {pages} {023601} (\bibinfo {year} {2018})}\BibitemShut
  {NoStop}%
\bibitem [{\citenamefont {Hazzard}\ \emph {et~al.}(2013)\citenamefont
  {Hazzard}, \citenamefont {Manmana}, \citenamefont {Foss-Feig},\ and\
  \citenamefont {Rey}}]{Hazzard2013}%
  \BibitemOpen
  \bibfield  {author} {\bibinfo {author} {\bibfnamefont {K.~R.}\ \bibnamefont
  {Hazzard}}, \bibinfo {author} {\bibfnamefont {S.~R.}\ \bibnamefont
  {Manmana}}, \bibinfo {author} {\bibfnamefont {M.}~\bibnamefont {Foss-Feig}},\
  and\ \bibinfo {author} {\bibfnamefont {A.~M.}\ \bibnamefont {Rey}},\
  }\bibfield  {title} {\bibinfo {title} {{Far-from-equilibrium quantum
  magnetism with ultracold polar molecules}},\ }\href
  {https://doi.org/10.1103/PhysRevLett.110.075301} {\bibfield  {journal}
  {\bibinfo  {journal} {Physical Review Letters}\ }\textbf {\bibinfo {volume}
  {110}},\ \bibinfo {pages} {075301} (\bibinfo {year} {2013})}\BibitemShut
  {NoStop}%
\bibitem [{\citenamefont {Pi{\~{n}}eiro~Orioli}\ \emph
  {et~al.}(2018)\citenamefont {Pi{\~{n}}eiro~Orioli}, \citenamefont {Signoles},
  \citenamefont {Wildhagen}, \citenamefont {G{\"{u}}nter}, \citenamefont
  {Berges}, \citenamefont {Whitlock},\ and\ \citenamefont
  {Weidem{\"{u}}ller}}]{Orioli2018}%
  \BibitemOpen
  \bibfield  {author} {\bibinfo {author} {\bibfnamefont {A.}~\bibnamefont
  {Pi{\~{n}}eiro~Orioli}}, \bibinfo {author} {\bibfnamefont {A.}~\bibnamefont
  {Signoles}}, \bibinfo {author} {\bibfnamefont {H.}~\bibnamefont {Wildhagen}},
  \bibinfo {author} {\bibfnamefont {G.}~\bibnamefont {G{\"{u}}nter}}, \bibinfo
  {author} {\bibfnamefont {J.}~\bibnamefont {Berges}}, \bibinfo {author}
  {\bibfnamefont {S.}~\bibnamefont {Whitlock}},\ and\ \bibinfo {author}
  {\bibfnamefont {M.}~\bibnamefont {Weidem{\"{u}}ller}},\ }\bibfield  {title}
  {\bibinfo {title} {{Relaxation of an isolated dipolar-interacting Rydberg
  quantum spin system}},\ }\href {http://arxiv.org/abs/1703.05957} {\bibfield
  {journal} {\bibinfo  {journal} {Physical Review Letters}\ }\textbf {\bibinfo
  {volume} {120}},\ \bibinfo {pages} {63601} (\bibinfo {year}
  {2018})}\BibitemShut {NoStop}%
\bibitem [{\citenamefont {Simon}\ \emph {et~al.}(2011)\citenamefont {Simon},
  \citenamefont {Bakr}, \citenamefont {Ma}, \citenamefont {Tai}, \citenamefont
  {Preiss},\ and\ \citenamefont {Greiner}}]{Simon2011}%
  \BibitemOpen
  \bibfield  {author} {\bibinfo {author} {\bibfnamefont {J.}~\bibnamefont
  {Simon}}, \bibinfo {author} {\bibfnamefont {W.~S.}\ \bibnamefont {Bakr}},
  \bibinfo {author} {\bibfnamefont {R.}~\bibnamefont {Ma}}, \bibinfo {author}
  {\bibfnamefont {M.~E.}\ \bibnamefont {Tai}}, \bibinfo {author} {\bibfnamefont
  {P.~M.}\ \bibnamefont {Preiss}},\ and\ \bibinfo {author} {\bibfnamefont
  {M.}~\bibnamefont {Greiner}},\ }\bibfield  {title} {\bibinfo {title}
  {{Quantum simulation of antiferromagnetic spin chains in an optical
  lattice}},\ }\href {https://doi.org/10.1038/nature09994} {\bibfield
  {journal} {\bibinfo  {journal} {Nature}\ }\textbf {\bibinfo {volume} {472}},\
  \bibinfo {pages} {307} (\bibinfo {year} {2011})},\ \Eprint
  {https://arxiv.org/abs/1103.1372} {arXiv:1103.1372} \BibitemShut {NoStop}%
\bibitem [{\citenamefont {Jin}\ \emph {et~al.}(2013)\citenamefont {Jin},
  \citenamefont {Hazzard}, \citenamefont {Moses}, \citenamefont {Gadway},
  \citenamefont {Yan}, \citenamefont {Covey}, \citenamefont {Rey},\ and\
  \citenamefont {Ye}}]{Yan2013}%
  \BibitemOpen
  \bibfield  {author} {\bibinfo {author} {\bibfnamefont {D.~S.}\ \bibnamefont
  {Jin}}, \bibinfo {author} {\bibfnamefont {K.~R.~A.}\ \bibnamefont {Hazzard}},
  \bibinfo {author} {\bibfnamefont {S.~A.}\ \bibnamefont {Moses}}, \bibinfo
  {author} {\bibfnamefont {B.}~\bibnamefont {Gadway}}, \bibinfo {author}
  {\bibfnamefont {B.}~\bibnamefont {Yan}}, \bibinfo {author} {\bibfnamefont
  {J.~P.}\ \bibnamefont {Covey}}, \bibinfo {author} {\bibfnamefont {A.~M.}\
  \bibnamefont {Rey}},\ and\ \bibinfo {author} {\bibfnamefont {J.}~\bibnamefont
  {Ye}},\ }\bibfield  {title} {\bibinfo {title} {{Observation of dipolar
  spin-exchange interactions with lattice-confined polar molecules}},\ }\href
  {https://doi.org/10.1038/nature12483} {\bibfield  {journal} {\bibinfo
  {journal} {Nature}\ }\textbf {\bibinfo {volume} {501}},\ \bibinfo {pages}
  {521} (\bibinfo {year} {2013})}\BibitemShut {NoStop}%
\bibitem [{\citenamefont {Labuhn}\ \emph {et~al.}(2016)\citenamefont {Labuhn},
  \citenamefont {Barredo}, \citenamefont {Ravets}, \citenamefont {{De
  L{\'{e}}s{\'{e}}leuc}}, \citenamefont {Macr{\`{i}}}, \citenamefont {Lahaye},\
  and\ \citenamefont {Browaeys}}]{Labuhn2016}%
  \BibitemOpen
  \bibfield  {author} {\bibinfo {author} {\bibfnamefont {H.}~\bibnamefont
  {Labuhn}}, \bibinfo {author} {\bibfnamefont {D.}~\bibnamefont {Barredo}},
  \bibinfo {author} {\bibfnamefont {S.}~\bibnamefont {Ravets}}, \bibinfo
  {author} {\bibfnamefont {S.}~\bibnamefont {{De L{\'{e}}s{\'{e}}leuc}}},
  \bibinfo {author} {\bibfnamefont {T.}~\bibnamefont {Macr{\`{i}}}}, \bibinfo
  {author} {\bibfnamefont {T.}~\bibnamefont {Lahaye}},\ and\ \bibinfo {author}
  {\bibfnamefont {A.}~\bibnamefont {Browaeys}},\ }\bibfield  {title} {\bibinfo
  {title} {{Tunable two-dimensional arrays of single Rydberg atoms for
  realizing quantum Ising models}},\ }\href
  {https://doi.org/10.1038/nature18274} {\bibfield  {journal} {\bibinfo
  {journal} {Nature}\ }\textbf {\bibinfo {volume} {534}},\ \bibinfo {pages}
  {667} (\bibinfo {year} {2016})}\BibitemShut {NoStop}%
\bibitem [{\citenamefont {Lepoutre}\ \emph {et~al.}(2019)\citenamefont
  {Lepoutre}, \citenamefont {Schachenmayer}, \citenamefont {Gabardos},
  \citenamefont {Zhu}, \citenamefont {Naylor}, \citenamefont {Mar{\'{e}}chal},
  \citenamefont {Gorceix}, \citenamefont {Rey}, \citenamefont {Vernac},\ and\
  \citenamefont {Laburthe-Tolra}}]{Lepoutre2019}%
  \BibitemOpen
  \bibfield  {author} {\bibinfo {author} {\bibfnamefont {S.}~\bibnamefont
  {Lepoutre}}, \bibinfo {author} {\bibfnamefont {J.}~\bibnamefont
  {Schachenmayer}}, \bibinfo {author} {\bibfnamefont {L.}~\bibnamefont
  {Gabardos}}, \bibinfo {author} {\bibfnamefont {B.}~\bibnamefont {Zhu}},
  \bibinfo {author} {\bibfnamefont {B.}~\bibnamefont {Naylor}}, \bibinfo
  {author} {\bibfnamefont {E.}~\bibnamefont {Mar{\'{e}}chal}}, \bibinfo
  {author} {\bibfnamefont {O.}~\bibnamefont {Gorceix}}, \bibinfo {author}
  {\bibfnamefont {A.~M.}\ \bibnamefont {Rey}}, \bibinfo {author} {\bibfnamefont
  {L.}~\bibnamefont {Vernac}},\ and\ \bibinfo {author} {\bibfnamefont
  {B.}~\bibnamefont {Laburthe-Tolra}},\ }\bibfield  {title} {\bibinfo {title}
  {{Out-of-equilibrium quantum magnetism and thermalization in a spin-3
  many-body dipolar lattice system}},\ }\href
  {https://doi.org/10.1038/s41467-019-09699-5} {\bibfield  {journal} {\bibinfo
  {journal} {Nature Communications}\ }\textbf {\bibinfo {volume} {10}},\
  \bibinfo {pages} {1714} (\bibinfo {year} {2019})}\BibitemShut {NoStop}%
\bibitem [{\citenamefont {Patscheider}\ \emph {et~al.}(2020)\citenamefont
  {Patscheider}, \citenamefont {Zhu}, \citenamefont {Chomaz}, \citenamefont
  {Petter}, \citenamefont {Baier}, \citenamefont {Rey}, \citenamefont
  {Ferlaino},\ and\ \citenamefont {Mark}}]{Patscheider2019}%
  \BibitemOpen
  \bibfield  {author} {\bibinfo {author} {\bibfnamefont {A.}~\bibnamefont
  {Patscheider}}, \bibinfo {author} {\bibfnamefont {B.}~\bibnamefont {Zhu}},
  \bibinfo {author} {\bibfnamefont {L.}~\bibnamefont {Chomaz}}, \bibinfo
  {author} {\bibfnamefont {D.}~\bibnamefont {Petter}}, \bibinfo {author}
  {\bibfnamefont {S.}~\bibnamefont {Baier}}, \bibinfo {author} {\bibfnamefont
  {A.-M.}\ \bibnamefont {Rey}}, \bibinfo {author} {\bibfnamefont
  {F.}~\bibnamefont {Ferlaino}},\ and\ \bibinfo {author} {\bibfnamefont
  {M.~J.}\ \bibnamefont {Mark}},\ }\bibfield  {title} {\bibinfo {title}
  {Controlling dipolar exchange interactions in a dense three-dimensional array
  of large-spin fermions},\ }\href
  {https://doi.org/10.1103/PhysRevResearch.2.023050} {\bibfield  {journal}
  {\bibinfo  {journal} {Phys. Rev. Research}\ }\textbf {\bibinfo {volume}
  {2}},\ \bibinfo {pages} {023050} (\bibinfo {year} {2020})}\BibitemShut
  {NoStop}%
\bibitem [{\citenamefont {Whitlock}\ \emph {et~al.}(2017)\citenamefont
  {Whitlock}, \citenamefont {Glaetzle},\ and\ \citenamefont
  {Hannaford}}]{Whitlock2017}%
  \BibitemOpen
  \bibfield  {author} {\bibinfo {author} {\bibfnamefont {S.}~\bibnamefont
  {Whitlock}}, \bibinfo {author} {\bibfnamefont {A.~W.}\ \bibnamefont
  {Glaetzle}},\ and\ \bibinfo {author} {\bibfnamefont {P.}~\bibnamefont
  {Hannaford}},\ }\bibfield  {title} {\bibinfo {title} {Simulating quantum spin
  models using rydberg-excited atomic ensembles in magnetic microtrap arrays},\
  }\href {https://doi.org/10.1088/1361-6455/aa6149} {\bibfield  {journal}
  {\bibinfo  {journal} {Journal of Physics B: Atomic, Molecular and Optical
  Physics}\ }\textbf {\bibinfo {volume} {50}},\ \bibinfo {pages} {074001}
  (\bibinfo {year} {2017})}\BibitemShut {NoStop}%
\bibitem [{\citenamefont {Nguyen}\ \emph {et~al.}(2018)\citenamefont {Nguyen},
  \citenamefont {Raimond}, \citenamefont {Sayrin}, \citenamefont {Corti\~nas},
  \citenamefont {Cantat-Moltrecht}, \citenamefont {Assemat}, \citenamefont
  {Dotsenko}, \citenamefont {Gleyzes}, \citenamefont {Haroche}, \citenamefont
  {Roux}, \citenamefont {Jolicoeur},\ and\ \citenamefont {Brune}}]{Nguyen2018}%
  \BibitemOpen
  \bibfield  {author} {\bibinfo {author} {\bibfnamefont {T.~L.}\ \bibnamefont
  {Nguyen}}, \bibinfo {author} {\bibfnamefont {J.~M.}\ \bibnamefont {Raimond}},
  \bibinfo {author} {\bibfnamefont {C.}~\bibnamefont {Sayrin}}, \bibinfo
  {author} {\bibfnamefont {R.}~\bibnamefont {Corti\~nas}}, \bibinfo {author}
  {\bibfnamefont {T.}~\bibnamefont {Cantat-Moltrecht}}, \bibinfo {author}
  {\bibfnamefont {F.}~\bibnamefont {Assemat}}, \bibinfo {author} {\bibfnamefont
  {I.}~\bibnamefont {Dotsenko}}, \bibinfo {author} {\bibfnamefont
  {S.}~\bibnamefont {Gleyzes}}, \bibinfo {author} {\bibfnamefont
  {S.}~\bibnamefont {Haroche}}, \bibinfo {author} {\bibfnamefont
  {G.}~\bibnamefont {Roux}}, \bibinfo {author} {\bibfnamefont {T.}~\bibnamefont
  {Jolicoeur}},\ and\ \bibinfo {author} {\bibfnamefont {M.}~\bibnamefont
  {Brune}},\ }\bibfield  {title} {\bibinfo {title} {Towards quantum simulation
  with circular rydberg atoms},\ }\href
  {https://doi.org/10.1103/PhysRevX.8.011032} {\bibfield  {journal} {\bibinfo
  {journal} {Phys. Rev. X}\ }\textbf {\bibinfo {volume} {8}},\ \bibinfo {pages}
  {011032} (\bibinfo {year} {2018})}\BibitemShut {NoStop}%
\bibitem [{\citenamefont {Ferreira-Cao}\ \emph {et~al.}(2020)\citenamefont
  {Ferreira-Cao}, \citenamefont {Gavryusev}, \citenamefont {Franz},
  \citenamefont {Alves}, \citenamefont {Signoles}, \citenamefont {Z{\"u}rn},\
  and\ \citenamefont {Weidem{\"u}ller}}]{ferreira2020depletion}%
  \BibitemOpen
  \bibfield  {author} {\bibinfo {author} {\bibfnamefont {M.}~\bibnamefont
  {Ferreira-Cao}}, \bibinfo {author} {\bibfnamefont {V.}~\bibnamefont
  {Gavryusev}}, \bibinfo {author} {\bibfnamefont {T.}~\bibnamefont {Franz}},
  \bibinfo {author} {\bibfnamefont {R.~F.}\ \bibnamefont {Alves}}, \bibinfo
  {author} {\bibfnamefont {A.}~\bibnamefont {Signoles}}, \bibinfo {author}
  {\bibfnamefont {G.}~\bibnamefont {Z{\"u}rn}},\ and\ \bibinfo {author}
  {\bibfnamefont {M.}~\bibnamefont {Weidem{\"u}ller}},\ }\bibfield  {title}
  {\bibinfo {title} {Depletion imaging of rydberg atoms in cold atomic gases},\
  }\href@noop {} {\bibfield  {journal} {\bibinfo  {journal} {Journal of Physics
  B: Atomic, Molecular and Optical Physics}\ }\textbf {\bibinfo {volume}
  {53}},\ \bibinfo {pages} {084004} (\bibinfo {year} {2020})}\BibitemShut
  {NoStop}%
\bibitem [{\citenamefont {{\v{S}}ibali{\'{c}}}\ \emph
  {et~al.}(2017)\citenamefont {{\v{S}}ibali{\'{c}}}, \citenamefont {Pritchard},
  \citenamefont {Adams},\ and\ \citenamefont {Weatherill}}]{Sibalic2017}%
  \BibitemOpen
  \bibfield  {author} {\bibinfo {author} {\bibfnamefont {N.}~\bibnamefont
  {{\v{S}}ibali{\'{c}}}}, \bibinfo {author} {\bibfnamefont {J.~D.}\
  \bibnamefont {Pritchard}}, \bibinfo {author} {\bibfnamefont {C.~S.}\
  \bibnamefont {Adams}},\ and\ \bibinfo {author} {\bibfnamefont {K.~J.}\
  \bibnamefont {Weatherill}},\ }\bibfield  {title} {\bibinfo {title} {{ARC: An
  open-source library for calculating properties of alkali Rydberg atoms}},\
  }\href {https://doi.org/10.1016/j.cpc.2017.06.015} {\bibfield  {journal}
  {\bibinfo  {journal} {Computer Physics Communications}\ }\textbf {\bibinfo
  {volume} {220}},\ \bibinfo {pages} {319} (\bibinfo {year}
  {2017})}\BibitemShut {NoStop}%
\bibitem [{\citenamefont {Carter}\ and\ \citenamefont
  {Martin}(2013)}]{Carter2013}%
  \BibitemOpen
  \bibfield  {author} {\bibinfo {author} {\bibfnamefont {J.~D.}\ \bibnamefont
  {Carter}}\ and\ \bibinfo {author} {\bibfnamefont {J.~D.~D.}\ \bibnamefont
  {Martin}},\ }\bibfield  {title} {\bibinfo {title} {Coherent manipulation of
  cold rydberg atoms near the surface of an atom chip},\ }\href
  {https://doi.org/10.1103/PhysRevA.88.043429} {\bibfield  {journal} {\bibinfo
  {journal} {Phys. Rev. A}\ }\textbf {\bibinfo {volume} {88}},\ \bibinfo
  {pages} {043429} (\bibinfo {year} {2013})}\BibitemShut {NoStop}%
\bibitem [{\citenamefont {Hermann-Avigliano}\ \emph {et~al.}(2014)\citenamefont
  {Hermann-Avigliano}, \citenamefont {Teixeira}, \citenamefont {Nguyen},
  \citenamefont {Cantat-Moltrecht}, \citenamefont {Nogues}, \citenamefont
  {Dotsenko}, \citenamefont {Gleyzes}, \citenamefont {Raimond}, \citenamefont
  {Haroche},\ and\ \citenamefont {Brune}}]{Hermann2014}%
  \BibitemOpen
  \bibfield  {author} {\bibinfo {author} {\bibfnamefont {C.}~\bibnamefont
  {Hermann-Avigliano}}, \bibinfo {author} {\bibfnamefont {R.~C.}\ \bibnamefont
  {Teixeira}}, \bibinfo {author} {\bibfnamefont {T.~L.}\ \bibnamefont
  {Nguyen}}, \bibinfo {author} {\bibfnamefont {T.}~\bibnamefont
  {Cantat-Moltrecht}}, \bibinfo {author} {\bibfnamefont {G.}~\bibnamefont
  {Nogues}}, \bibinfo {author} {\bibfnamefont {I.}~\bibnamefont {Dotsenko}},
  \bibinfo {author} {\bibfnamefont {S.}~\bibnamefont {Gleyzes}}, \bibinfo
  {author} {\bibfnamefont {J.~M.}\ \bibnamefont {Raimond}}, \bibinfo {author}
  {\bibfnamefont {S.}~\bibnamefont {Haroche}},\ and\ \bibinfo {author}
  {\bibfnamefont {M.}~\bibnamefont {Brune}},\ }\bibfield  {title} {\bibinfo
  {title} {Long coherence times for rydberg qubits on a superconducting atom
  chip},\ }\href {https://doi.org/10.1103/PhysRevA.90.040502} {\bibfield
  {journal} {\bibinfo  {journal} {Phys. Rev. A}\ }\textbf {\bibinfo {volume}
  {90}},\ \bibinfo {pages} {040502} (\bibinfo {year} {2014})}\BibitemShut
  {NoStop}%
\bibitem [{\citenamefont {Hertz}(1909)}]{Hertz1909}%
  \BibitemOpen
  \bibfield  {author} {\bibinfo {author} {\bibfnamefont {P.}~\bibnamefont
  {Hertz}},\ }\bibfield  {title} {\bibinfo {title} {{\"Uber den gegenseitigen
  durchschnittlichen Abstand von Punkten, die mit bekannter mittlerer Dichte im
  Raume angeordnet sind}},\ }\href {https://doi.org/10.1007/BF01450410}
  {\bibfield  {journal} {\bibinfo  {journal} {Mathematische Annalen}\ }\textbf
  {\bibinfo {volume} {67}},\ \bibinfo {pages} {387} (\bibinfo {year}
  {1909})}\BibitemShut {NoStop}%
\bibitem [{\citenamefont {Phillips}(1996)}]{Phillips1996}%
  \BibitemOpen
  \bibfield  {author} {\bibinfo {author} {\bibfnamefont {J.~C.}\ \bibnamefont
  {Phillips}},\ }\bibfield  {title} {\bibinfo {title} {{Stretched exponential
  relaxation in molecular and electronic glasses}},\ }\href
  {https://doi.org/10.1088/0034-4885/59/9/003} {\bibfield  {journal} {\bibinfo
  {journal} {Reports on Progress in Physics}\ }\textbf {\bibinfo {volume}
  {59}},\ \bibinfo {pages} {1133} (\bibinfo {year} {1996})}\BibitemShut
  {NoStop}%
\bibitem [{\citenamefont {Fischer}\ \emph {et~al.}(2016)\citenamefont
  {Fischer}, \citenamefont {Maksymenko},\ and\ \citenamefont
  {Altman}}]{Fischer2016}%
  \BibitemOpen
  \bibfield  {author} {\bibinfo {author} {\bibfnamefont {M.~H.}\ \bibnamefont
  {Fischer}}, \bibinfo {author} {\bibfnamefont {M.}~\bibnamefont
  {Maksymenko}},\ and\ \bibinfo {author} {\bibfnamefont {E.}~\bibnamefont
  {Altman}},\ }\bibfield  {title} {\bibinfo {title} {{Dynamics of a
  Many-Body-Localized System Coupled to a Bath}},\ }\href
  {https://doi.org/10.1103/PhysRevLett.116.160401} {\bibfield  {journal}
  {\bibinfo  {journal} {Physical Review Letters}\ }\textbf {\bibinfo {volume}
  {116}},\ \bibinfo {pages} {160401} (\bibinfo {year} {2016})}\BibitemShut
  {NoStop}%
\bibitem [{\citenamefont {Mukherjee}\ \emph {et~al.}(2016)\citenamefont
  {Mukherjee}, \citenamefont {Killian},\ and\ \citenamefont
  {Hazzard}}]{Mukherjee2016}%
  \BibitemOpen
  \bibfield  {author} {\bibinfo {author} {\bibfnamefont {R.}~\bibnamefont
  {Mukherjee}}, \bibinfo {author} {\bibfnamefont {T.~C.}\ \bibnamefont
  {Killian}},\ and\ \bibinfo {author} {\bibfnamefont {K.~R.~A.}\ \bibnamefont
  {Hazzard}},\ }\bibfield  {title} {\bibinfo {title} {Accessing rydberg-dressed
  interactions using many-body ramsey dynamics},\ }\href
  {https://doi.org/10.1103/PhysRevA.94.053422} {\bibfield  {journal} {\bibinfo
  {journal} {Phys. Rev. A}\ }\textbf {\bibinfo {volume} {94}},\ \bibinfo
  {pages} {053422} (\bibinfo {year} {2016})}\BibitemShut {NoStop}%
\bibitem [{\citenamefont {Schau{\ss}}\ \emph {et~al.}(2012)\citenamefont
  {Schau{\ss}}, \citenamefont {Cheneau}, \citenamefont {Endres}, \citenamefont
  {Fukuhara}, \citenamefont {Hild}, \citenamefont {Omran}, \citenamefont
  {Pohl}, \citenamefont {Gross}, \citenamefont {Kuhr},\ and\ \citenamefont
  {Bloch}}]{Schauss2012}%
  \BibitemOpen
  \bibfield  {author} {\bibinfo {author} {\bibfnamefont {P.}~\bibnamefont
  {Schau{\ss}}}, \bibinfo {author} {\bibfnamefont {M.}~\bibnamefont {Cheneau}},
  \bibinfo {author} {\bibfnamefont {M.}~\bibnamefont {Endres}}, \bibinfo
  {author} {\bibfnamefont {T.}~\bibnamefont {Fukuhara}}, \bibinfo {author}
  {\bibfnamefont {S.}~\bibnamefont {Hild}}, \bibinfo {author} {\bibfnamefont
  {A.}~\bibnamefont {Omran}}, \bibinfo {author} {\bibfnamefont
  {T.}~\bibnamefont {Pohl}}, \bibinfo {author} {\bibfnamefont {C.}~\bibnamefont
  {Gross}}, \bibinfo {author} {\bibfnamefont {S.}~\bibnamefont {Kuhr}},\ and\
  \bibinfo {author} {\bibfnamefont {I.}~\bibnamefont {Bloch}},\ }\bibfield
  {title} {\bibinfo {title} {{Observation of mesoscopic crystalline structures
  in a two-dimensional Rydberg gas}},\ }\href
  {https://doi.org/10.1038/nature11596} {\bibfield  {journal} {\bibinfo
  {journal} {Nature}\ ,\ \bibinfo {pages} {87}} (\bibinfo {year} {2012})},\
  \Eprint {https://arxiv.org/abs/1209.0944} {arXiv:1209.0944} \BibitemShut
  {NoStop}%
\bibitem [{\citenamefont {Bettelli}\ \emph {et~al.}(2013)\citenamefont
  {Bettelli}, \citenamefont {Maxwell}, \citenamefont {Fernholz}, \citenamefont
  {Adams}, \citenamefont {Lesanovsky},\ and\ \citenamefont
  {Ates}}]{Bettelli2013}%
  \BibitemOpen
  \bibfield  {author} {\bibinfo {author} {\bibfnamefont {S.}~\bibnamefont
  {Bettelli}}, \bibinfo {author} {\bibfnamefont {D.}~\bibnamefont {Maxwell}},
  \bibinfo {author} {\bibfnamefont {T.}~\bibnamefont {Fernholz}}, \bibinfo
  {author} {\bibfnamefont {C.~S.}\ \bibnamefont {Adams}}, \bibinfo {author}
  {\bibfnamefont {I.}~\bibnamefont {Lesanovsky}},\ and\ \bibinfo {author}
  {\bibfnamefont {C.}~\bibnamefont {Ates}},\ }\bibfield  {title} {\bibinfo
  {title} {Exciton dynamics in emergent rydberg lattices},\ }\href
  {https://doi.org/10.1103/PhysRevA.88.043436} {\bibfield  {journal} {\bibinfo
  {journal} {Phys. Rev. A}\ }\textbf {\bibinfo {volume} {88}},\ \bibinfo
  {pages} {043436} (\bibinfo {year} {2013})}\BibitemShut {NoStop}%
\bibitem [{\citenamefont {Urvoy}\ \emph {et~al.}(2015)\citenamefont {Urvoy},
  \citenamefont {Ripka}, \citenamefont {Lesanovsky}, \citenamefont {Booth},
  \citenamefont {Shaffer}, \citenamefont {Pfau},\ and\ \citenamefont
  {L{\"{o}}w}}]{Urvoy2015}%
  \BibitemOpen
  \bibfield  {author} {\bibinfo {author} {\bibfnamefont {A.}~\bibnamefont
  {Urvoy}}, \bibinfo {author} {\bibfnamefont {F.}~\bibnamefont {Ripka}},
  \bibinfo {author} {\bibfnamefont {I.}~\bibnamefont {Lesanovsky}}, \bibinfo
  {author} {\bibfnamefont {D.}~\bibnamefont {Booth}}, \bibinfo {author}
  {\bibfnamefont {J.~P.}\ \bibnamefont {Shaffer}}, \bibinfo {author}
  {\bibfnamefont {T.}~\bibnamefont {Pfau}},\ and\ \bibinfo {author}
  {\bibfnamefont {R.}~\bibnamefont {L{\"{o}}w}},\ }\bibfield  {title} {\bibinfo
  {title} {{Strongly Correlated Growth of Rydberg Aggregates in a Vapor
  Cell}},\ }\href {https://doi.org/10.1103/PhysRevLett.114.203002} {\bibfield
  {journal} {\bibinfo  {journal} {Physical Review Letters}\ }\textbf {\bibinfo
  {volume} {114}},\ \bibinfo {pages} {203002} (\bibinfo {year}
  {2015})}\BibitemShut {NoStop}%
\bibitem [{\citenamefont {Polkovnikov}(2010)}]{Polkovnikov2010}%
  \BibitemOpen
  \bibfield  {author} {\bibinfo {author} {\bibfnamefont {A.}~\bibnamefont
  {Polkovnikov}},\ }\bibfield  {title} {\bibinfo {title} {{Phase space
  representation of quantum dynamics}},\ }\href
  {https://doi.org/10.1016/j.aop.2010.02.006} {\bibfield  {journal} {\bibinfo
  {journal} {Annals of Physics}\ }\textbf {\bibinfo {volume} {325}},\ \bibinfo
  {pages} {1790} (\bibinfo {year} {2010})}\BibitemShut {NoStop}%
\bibitem [{\citenamefont {Schachenmayer}\ \emph {et~al.}(2015)\citenamefont
  {Schachenmayer}, \citenamefont {Pikovski},\ and\ \citenamefont
  {Rey}}]{Schachenmayer2015}%
  \BibitemOpen
  \bibfield  {author} {\bibinfo {author} {\bibfnamefont {J.}~\bibnamefont
  {Schachenmayer}}, \bibinfo {author} {\bibfnamefont {A.}~\bibnamefont
  {Pikovski}},\ and\ \bibinfo {author} {\bibfnamefont {A.~M.}\ \bibnamefont
  {Rey}},\ }\bibfield  {title} {\bibinfo {title} {{Many-body quantum spin
  dynamics with monte carlo trajectories on a discrete phase space}},\ }\href
  {https://doi.org/10.1103/PhysRevX.5.011022} {\bibfield  {journal} {\bibinfo
  {journal} {Physical Review X}\ }\textbf {\bibinfo {volume} {5}},\ \bibinfo
  {pages} {011022} (\bibinfo {year} {2015})}\BibitemShut {NoStop}%
\bibitem [{\citenamefont {Emch}(1966)}]{Emch1966}%
  \BibitemOpen
  \bibfield  {author} {\bibinfo {author} {\bibfnamefont {G.~G.}\ \bibnamefont
  {Emch}},\ }\bibfield  {title} {\bibinfo {title} {{Non-Markovian Model for the
  Approach to Equilibrium}}\ }\href {https://doi.org/10.1063/1.1705023}
  {10.1063/1.1705023} (\bibinfo {year} {1966})\BibitemShut {NoStop}%
\bibitem [{\citenamefont {Radin}(1970)}]{Radin1970}%
  \BibitemOpen
  \bibfield  {author} {\bibinfo {author} {\bibfnamefont {C.}~\bibnamefont
  {Radin}},\ }\bibfield  {title} {\bibinfo {title} {{Approach to equilibrium in
  a simple model}},\ }\href {https://doi.org/10.1063/1.1665079} {\bibfield
  {journal} {\bibinfo  {journal} {Journal of Mathematical Physics}\ }\textbf
  {\bibinfo {volume} {11}},\ \bibinfo {pages} {2945} (\bibinfo {year}
  {1970})}\BibitemShut {NoStop}%
\bibitem [{\citenamefont {Takei}\ \emph {et~al.}(2016)\citenamefont {Takei},
  \citenamefont {Sommer}, \citenamefont {Genes}, \citenamefont {Pupillo},
  \citenamefont {Goto}, \citenamefont {Koyasu}, \citenamefont {Chiba},
  \citenamefont {Weidem{\"{u}}ller},\ and\ \citenamefont {Ohmori}}]{Takei2016}%
  \BibitemOpen
  \bibfield  {author} {\bibinfo {author} {\bibfnamefont {N.}~\bibnamefont
  {Takei}}, \bibinfo {author} {\bibfnamefont {C.}~\bibnamefont {Sommer}},
  \bibinfo {author} {\bibfnamefont {C.}~\bibnamefont {Genes}}, \bibinfo
  {author} {\bibfnamefont {G.}~\bibnamefont {Pupillo}}, \bibinfo {author}
  {\bibfnamefont {H.}~\bibnamefont {Goto}}, \bibinfo {author} {\bibfnamefont
  {K.}~\bibnamefont {Koyasu}}, \bibinfo {author} {\bibfnamefont
  {H.}~\bibnamefont {Chiba}}, \bibinfo {author} {\bibfnamefont
  {M.}~\bibnamefont {Weidem{\"{u}}ller}},\ and\ \bibinfo {author}
  {\bibfnamefont {K.}~\bibnamefont {Ohmori}},\ }\bibfield  {title} {\bibinfo
  {title} {{Direct observation of ultrafast many-body electron dynamics in an
  ultracold Rydberg gas}},\ }\href {https://doi.org/10.1038/ncomms13449}
  {\bibfield  {journal} {\bibinfo  {journal} {Nature Communications}\ }\textbf
  {\bibinfo {volume} {7}},\ \bibinfo {pages} {1} (\bibinfo {year} {2016})},\
  \Eprint {https://arxiv.org/abs/1504.03635} {arXiv:1504.03635} \BibitemShut
  {NoStop}%
\bibitem [{\citenamefont {Sommer}\ \emph {et~al.}(2016)\citenamefont {Sommer},
  \citenamefont {Pupillo}, \citenamefont {Takei}, \citenamefont {Takeda},
  \citenamefont {Tanaka}, \citenamefont {Ohmori},\ and\ \citenamefont
  {Genes}}]{Sommer2016}%
  \BibitemOpen
  \bibfield  {author} {\bibinfo {author} {\bibfnamefont {C.}~\bibnamefont
  {Sommer}}, \bibinfo {author} {\bibfnamefont {G.}~\bibnamefont {Pupillo}},
  \bibinfo {author} {\bibfnamefont {N.}~\bibnamefont {Takei}}, \bibinfo
  {author} {\bibfnamefont {S.}~\bibnamefont {Takeda}}, \bibinfo {author}
  {\bibfnamefont {A.}~\bibnamefont {Tanaka}}, \bibinfo {author} {\bibfnamefont
  {K.}~\bibnamefont {Ohmori}},\ and\ \bibinfo {author} {\bibfnamefont
  {C.}~\bibnamefont {Genes}},\ }\bibfield  {title} {\bibinfo {title}
  {Time-domain ramsey interferometry with interacting rydberg atoms},\ }\href
  {https://doi.org/10.1103/PhysRevA.94.053607} {\bibfield  {journal} {\bibinfo
  {journal} {Phys. Rev. A}\ }\textbf {\bibinfo {volume} {94}},\ \bibinfo
  {pages} {053607} (\bibinfo {year} {2016})}\BibitemShut {NoStop}%
\bibitem [{\citenamefont {Everest}\ \emph {et~al.}(2017)\citenamefont
  {Everest}, \citenamefont {Lesanovsky}, \citenamefont {Garrahan},\ and\
  \citenamefont {Levi}}]{Everest2017}%
  \BibitemOpen
  \bibfield  {author} {\bibinfo {author} {\bibfnamefont {B.}~\bibnamefont
  {Everest}}, \bibinfo {author} {\bibfnamefont {I.}~\bibnamefont {Lesanovsky}},
  \bibinfo {author} {\bibfnamefont {J.~P.}\ \bibnamefont {Garrahan}},\ and\
  \bibinfo {author} {\bibfnamefont {E.}~\bibnamefont {Levi}},\ }\bibfield
  {title} {\bibinfo {title} {Role of interactions in a dissipative many-body
  localized system},\ }\href {https://doi.org/10.1103/PhysRevB.95.024310}
  {\bibfield  {journal} {\bibinfo  {journal} {Phys. Rev. B}\ }\textbf {\bibinfo
  {volume} {95}},\ \bibinfo {pages} {024310} (\bibinfo {year}
  {2017})}\BibitemShut {NoStop}%
\bibitem [{\citenamefont {Bouchaud}\ \emph {et~al.}(1997)\citenamefont
  {Bouchaud}, \citenamefont {Cugliandolo}, \citenamefont {Kurchan},\ and\
  \citenamefont {Mezard}}]{Bouchaud1998}%
  \BibitemOpen
  \bibfield  {author} {\bibinfo {author} {\bibfnamefont {J.-P.}\ \bibnamefont
  {Bouchaud}}, \bibinfo {author} {\bibfnamefont {L.~F.}\ \bibnamefont
  {Cugliandolo}}, \bibinfo {author} {\bibfnamefont {J.}~\bibnamefont
  {Kurchan}},\ and\ \bibinfo {author} {\bibfnamefont {M.}~\bibnamefont
  {Mezard}},\ }\bibfield  {title} {\bibinfo {title} {{Out of equilibrium
  dynamics in spin-glasses and other glassy systems}},\ }\href
  {http://arxiv.org/abs/cond-mat/9702070} {\bibfield  {journal} {\bibinfo
  {journal} {Spin glasses and random fields}\ ,\ \bibinfo {pages} {161}}
  (\bibinfo {year} {1997})},\ \Eprint {https://arxiv.org/abs/9702070}
  {arXiv:9702070 [cond-mat]} \BibitemShut {NoStop}%
\bibitem [{\citenamefont {{De Dominicis}}\ \emph {et~al.}(1985)\citenamefont
  {{De Dominicis}}, \citenamefont {Orland},\ and\ \citenamefont
  {Lain{\'{e}}e}}]{DeDominicis1985}%
  \BibitemOpen
  \bibfield  {author} {\bibinfo {author} {\bibfnamefont {C.}~\bibnamefont {{De
  Dominicis}}}, \bibinfo {author} {\bibfnamefont {H.}~\bibnamefont {Orland}},\
  and\ \bibinfo {author} {\bibfnamefont {F.}~\bibnamefont {Lain{\'{e}}e}},\
  }\bibfield  {title} {\bibinfo {title} {{Stretched exponential relaxation in
  systems with random free energies}},\ }\href
  {https://doi.org/10.1051/jphyslet:019850046011046300} {\bibfield  {journal}
  {\bibinfo  {journal} {Journal de Physique Lettres}\ }\textbf {\bibinfo
  {volume} {46}},\ \bibinfo {pages} {463} (\bibinfo {year} {1985})}\BibitemShut
  {NoStop}%
\bibitem [{\citenamefont {Young}\ and\ \citenamefont
  {Rieger}(1996)}]{young1996numerical}%
  \BibitemOpen
  \bibfield  {author} {\bibinfo {author} {\bibfnamefont {A.}~\bibnamefont
  {Young}}\ and\ \bibinfo {author} {\bibfnamefont {H.}~\bibnamefont {Rieger}},\
  }\bibfield  {title} {\bibinfo {title} {Numerical study of the random
  transverse-field ising spin chain},\ }\href@noop {} {\bibfield  {journal}
  {\bibinfo  {journal} {Physical Review B}\ }\textbf {\bibinfo {volume} {53}},\
  \bibinfo {pages} {8486} (\bibinfo {year} {1996})}\BibitemShut {NoStop}%
\bibitem [{\citenamefont {Crespo}\ \emph {et~al.}(2013)\citenamefont {Crespo},
  \citenamefont {Andreanov},\ and\ \citenamefont {Seriani}}]{Crespo2013}%
  \BibitemOpen
  \bibfield  {author} {\bibinfo {author} {\bibfnamefont {Y.}~\bibnamefont
  {Crespo}}, \bibinfo {author} {\bibfnamefont {A.}~\bibnamefont {Andreanov}},\
  and\ \bibinfo {author} {\bibfnamefont {N.}~\bibnamefont {Seriani}},\
  }\bibfield  {title} {\bibinfo {title} {Competing antiferromagnetic and
  spin-glass phases in a hollandite structure},\ }\href
  {https://doi.org/10.1103/PhysRevB.88.014202} {\bibfield  {journal} {\bibinfo
  {journal} {Phys. Rev. B}\ }\textbf {\bibinfo {volume} {88}},\ \bibinfo
  {pages} {014202} (\bibinfo {year} {2013})}\BibitemShut {NoStop}%
\bibitem [{\citenamefont {Leviatan}\ \emph {et~al.}(2017)\citenamefont
  {Leviatan}, \citenamefont {Pollmann}, \citenamefont {Bardarson},
  \citenamefont {Huse},\ and\ \citenamefont {Altman}}]{Leviatan2017}%
  \BibitemOpen
  \bibfield  {author} {\bibinfo {author} {\bibfnamefont {E.}~\bibnamefont
  {Leviatan}}, \bibinfo {author} {\bibfnamefont {F.}~\bibnamefont {Pollmann}},
  \bibinfo {author} {\bibfnamefont {J.~H.}\ \bibnamefont {Bardarson}}, \bibinfo
  {author} {\bibfnamefont {D.~A.}\ \bibnamefont {Huse}},\ and\ \bibinfo
  {author} {\bibfnamefont {E.}~\bibnamefont {Altman}},\ }\bibfield  {title}
  {\bibinfo {title} {{Quantum thermalization dynamics with Matrix-Product
  States}},\ }\href {http://arxiv.org/abs/1702.08894} {\  (\bibinfo {year}
  {2017})},\ \Eprint {https://arxiv.org/abs/1702.08894} {arXiv:1702.08894}
  \BibitemShut {NoStop}%
\bibitem [{\citenamefont {Igl{\'{o}}i}\ and\ \citenamefont
  {Monthus}(2018)}]{Igloi2018}%
  \BibitemOpen
  \bibfield  {author} {\bibinfo {author} {\bibfnamefont {F.}~\bibnamefont
  {Igl{\'{o}}i}}\ and\ \bibinfo {author} {\bibfnamefont {C.}~\bibnamefont
  {Monthus}},\ }\bibfield  {title} {\bibinfo {title} {{Strong disorder RG
  approach – a short review of recent developments}},\ }\href
  {https://doi.org/10.1140/epjb/e2018-90434-8} {\bibfield  {journal} {\bibinfo
  {journal} {The European Physical Journal B}\ }\textbf {\bibinfo {volume}
  {91}},\ \bibinfo {pages} {290} (\bibinfo {year} {2018})}\BibitemShut
  {NoStop}%
\bibitem [{\citenamefont {H{\'{e}}risson}\ and\ \citenamefont
  {Ocio}(2002)}]{Herisson2002}%
  \BibitemOpen
  \bibfield  {author} {\bibinfo {author} {\bibfnamefont {D.}~\bibnamefont
  {H{\'{e}}risson}}\ and\ \bibinfo {author} {\bibfnamefont {M.}~\bibnamefont
  {Ocio}},\ }\bibfield  {title} {\bibinfo {title} {{Fluctuation-Dissipation
  Ratio of a Spin Glass in the Aging Regime}},\ }\href
  {https://doi.org/10.1103/PhysRevLett.88.257202} {\bibfield  {journal}
  {\bibinfo  {journal} {Physical Review Letters}\ }\textbf {\bibinfo {volume}
  {88}},\ \bibinfo {pages} {257202} (\bibinfo {year} {2002})}\BibitemShut
  {NoStop}%
\bibitem [{\citenamefont {Anderson}(1988)}]{Anderson1988}%
  \BibitemOpen
  \bibfield  {author} {\bibinfo {author} {\bibfnamefont {P.~W.}\ \bibnamefont
  {Anderson}},\ }\bibfield  {title} {\bibinfo {title} {{Spin Glass II: Is There
  a Phase Transition?}},\ }\href {https://doi.org/10.1063/1.2811336} {\bibfield
   {journal} {\bibinfo  {journal} {Physics Today}\ }\textbf {\bibinfo {volume}
  {41}},\ \bibinfo {pages} {9} (\bibinfo {year} {1988})}\BibitemShut {NoStop}%
\bibitem [{\citenamefont {Edelstein}\ and\ \citenamefont
  {Gallagher}(1979)}]{gallagher_1994}%
  \BibitemOpen
  \bibfield  {author} {\bibinfo {author} {\bibfnamefont {S.~A.}\ \bibnamefont
  {Edelstein}}\ and\ \bibinfo {author} {\bibfnamefont {T.~F.}\ \bibnamefont
  {Gallagher}},\ }\href {https://doi.org/10.1016/S0065-2199(08)60132-3} {\emph
  {\bibinfo {title} {Advances in Atomic and Molecular Physics}}},\ \bibinfo
  {series} {Cambridge Monographs on Atomic, Molecular and Chemical Physics},
  Vol.~\bibinfo {volume} {14}\ (\bibinfo  {publisher} {Cambridge University
  Press},\ \bibinfo {address} {Cambridge},\ \bibinfo {year} {1979})\ pp.\
  \bibinfo {pages} {365--392}\BibitemShut {NoStop}%
\bibitem [{\citenamefont {Tsitouras}(2011)}]{Tsitouras2011}%
  \BibitemOpen
  \bibfield  {author} {\bibinfo {author} {\bibfnamefont {C.}~\bibnamefont
  {Tsitouras}},\ }\bibfield  {title} {\bibinfo {title} {{Runge–Kutta pairs of
  order 5(4) satisfying only the first column simplifying assumption}},\ }\href
  {https://doi.org/10.1016/J.CAMWA.2011.06.002} {\bibfield  {journal} {\bibinfo
   {journal} {Computers {\&} Mathematics with Applications}\ }\textbf {\bibinfo
  {volume} {62}},\ \bibinfo {pages} {770} (\bibinfo {year} {2011})}\BibitemShut
  {NoStop}%
\bibitem [{\citenamefont {Rackauckas}\ and\ \citenamefont
  {Nie}(2017)}]{Rackauckas2017}%
  \BibitemOpen
  \bibfield  {author} {\bibinfo {author} {\bibfnamefont {C.}~\bibnamefont
  {Rackauckas}}\ and\ \bibinfo {author} {\bibfnamefont {Q.}~\bibnamefont
  {Nie}},\ }\bibfield  {title} {\bibinfo {title} {{DifferentialEquations.jl –
  A Performant and Feature-Rich Ecosystem for Solving Differential Equations in
  Julia}},\ }\bibfield  {journal} {\bibinfo  {journal} {Journal of Open
  Research Software}\ }\textbf {\bibinfo {volume} {5}},\ \href
  {https://doi.org/10.5334/jors.151} {10.5334/jors.151} (\bibinfo {year}
  {2017})\BibitemShut {NoStop}%
\bibitem [{\citenamefont {Hazzard}\ \emph {et~al.}(2014)\citenamefont
  {Hazzard}, \citenamefont {Gadway}, \citenamefont {Foss-Feig}, \citenamefont
  {Yan}, \citenamefont {Moses}, \citenamefont {Covey}, \citenamefont {Yao},
  \citenamefont {Lukin}, \citenamefont {Ye}, \citenamefont {Jin},\ and\
  \citenamefont {Rey}}]{Hazzard2014}%
  \BibitemOpen
  \bibfield  {author} {\bibinfo {author} {\bibfnamefont {K.~R.}\ \bibnamefont
  {Hazzard}}, \bibinfo {author} {\bibfnamefont {B.}~\bibnamefont {Gadway}},
  \bibinfo {author} {\bibfnamefont {M.}~\bibnamefont {Foss-Feig}}, \bibinfo
  {author} {\bibfnamefont {B.}~\bibnamefont {Yan}}, \bibinfo {author}
  {\bibfnamefont {S.~A.}\ \bibnamefont {Moses}}, \bibinfo {author}
  {\bibfnamefont {J.~P.}\ \bibnamefont {Covey}}, \bibinfo {author}
  {\bibfnamefont {N.~Y.}\ \bibnamefont {Yao}}, \bibinfo {author} {\bibfnamefont
  {M.~D.}\ \bibnamefont {Lukin}}, \bibinfo {author} {\bibfnamefont
  {J.}~\bibnamefont {Ye}}, \bibinfo {author} {\bibfnamefont {D.~S.}\
  \bibnamefont {Jin}},\ and\ \bibinfo {author} {\bibfnamefont {A.~M.}\
  \bibnamefont {Rey}},\ }\bibfield  {title} {\bibinfo {title} {{Many-body
  dynamics of dipolar molecules in an optical lattice}},\ }\href
  {https://doi.org/10.1103/PhysRevLett.113.195302} {\bibfield  {journal}
  {\bibinfo  {journal} {Physical Review Letters}\ }\textbf {\bibinfo {volume}
  {113}},\ \bibinfo {pages} {195302} (\bibinfo {year} {2014})}\BibitemShut
  {NoStop}%
\bibitem [{\citenamefont {Kohlrausch}(1854)}]{Kohlrausch1854}%
  \BibitemOpen
  \bibfield  {author} {\bibinfo {author} {\bibfnamefont {R.}~\bibnamefont
  {Kohlrausch}},\ }\bibfield  {title} {\bibinfo {title} {{Theorie des
  elektrischen R{\"{u}}ckstandes in der Leidener Flasche}},\ }\href
  {https://doi.org/10.1002/andp.18541670103} {\bibfield  {journal} {\bibinfo
  {journal} {Annalen der Physik}\ }\textbf {\bibinfo {volume} {167}},\ \bibinfo
  {pages} {56} (\bibinfo {year} {1854})}\BibitemShut {NoStop}%
\bibitem [{\citenamefont {Weimer}\ \emph {et~al.}(2008)\citenamefont {Weimer},
  \citenamefont {L{\"{o}}w}, \citenamefont {Pfau},\ and\ \citenamefont
  {B{\"{u}}chler}}]{Weimer2008}%
  \BibitemOpen
  \bibfield  {author} {\bibinfo {author} {\bibfnamefont {H.}~\bibnamefont
  {Weimer}}, \bibinfo {author} {\bibfnamefont {R.}~\bibnamefont {L{\"{o}}w}},
  \bibinfo {author} {\bibfnamefont {T.}~\bibnamefont {Pfau}},\ and\ \bibinfo
  {author} {\bibfnamefont {H.~P.}\ \bibnamefont {B{\"{u}}chler}},\ }\bibfield
  {title} {\bibinfo {title} {{Quantum critical behavior in strongly interacting
  rydberg gases}},\ }\href {https://doi.org/10.1103/PhysRevLett.101.250601}
  {\bibfield  {journal} {\bibinfo  {journal} {Physical Review Letters}\
  }\textbf {\bibinfo {volume} {101}},\ \bibinfo {pages} {250601} (\bibinfo
  {year} {2008})},\ \Eprint {https://arxiv.org/abs/0806.3754} {arXiv:0806.3754}
  \BibitemShut {NoStop}%
\bibitem [{\citenamefont {G{\"{a}}rttner}\ \emph {et~al.}(2012)\citenamefont
  {G{\"{a}}rttner}, \citenamefont {Heeg}, \citenamefont {Gasenzer},\ and\
  \citenamefont {Evers}}]{Garttner2012}%
  \BibitemOpen
  \bibfield  {author} {\bibinfo {author} {\bibfnamefont {M.}~\bibnamefont
  {G{\"{a}}rttner}}, \bibinfo {author} {\bibfnamefont {K.~P.}\ \bibnamefont
  {Heeg}}, \bibinfo {author} {\bibfnamefont {T.}~\bibnamefont {Gasenzer}},\
  and\ \bibinfo {author} {\bibfnamefont {J.}~\bibnamefont {Evers}},\ }\bibfield
   {title} {\bibinfo {title} {{Finite-size effects in strongly interacting
  Rydberg gases}},\ }\href {https://doi.org/10.1103/PhysRevA.86.033422}
  {\bibfield  {journal} {\bibinfo  {journal} {Physical Review A - Atomic,
  Molecular, and Optical Physics}\ }\textbf {\bibinfo {volume} {86}},\ \bibinfo
  {pages} {033422} (\bibinfo {year} {2012})}\BibitemShut {NoStop}%
\bibitem [{\citenamefont {Kullback}\ and\ \citenamefont
  {Leibler}(2007)}]{Kullback1951}%
  \BibitemOpen
  \bibfield  {author} {\bibinfo {author} {\bibfnamefont {S.}~\bibnamefont
  {Kullback}}\ and\ \bibinfo {author} {\bibfnamefont {R.~A.}\ \bibnamefont
  {Leibler}},\ }\bibfield  {title} {\bibinfo {title} {{On Information and
  Sufficiency}},\ }\href {https://doi.org/10.1214/aoms/1177729694} {\bibfield
  {journal} {\bibinfo  {journal} {The Annals of Mathematical Statistics}\
  }\textbf {\bibinfo {volume} {22}},\ \bibinfo {pages} {79} (\bibinfo {year}
  {2007})}\BibitemShut {NoStop}%
\end{thebibliography}%

\end{document}